\documentclass{aa}
\usepackage{graphics}
\begin{document}
\thesaurus{02.01.1; 02.18.5; 02.18.7; 08.15.9 Cas~A; 13.18.5}
\title{ On energy-dependent propagation effects and acceleration 
sites of relativistic electrons in Cassiopeia~A}
\author{ A. M. Atoyan \inst{1,2}
\and R. J. Tuffs \inst1
\and F. A. Aharonian \inst1
\and H. J. V\"olk \inst1}
\offprints{Richard.Tuffs@mpi-hd.mpg.de}
\institute{Max Planck Institut f\"ur Kernphysik, Saupfercheckweg 1,
           D--69117 Heidelberg, Germany
\and Yerevan Physics Institute, 375036 Yerevan, Armenia} 
\date{Received 24 December 1998 / accepted 5 July 1999}
\titlerunning{Energy-dependent propagation effects in Cas A} 
\maketitle

\begin{abstract}
We consider the effect of energy dependent propagation of relativistic 
electrons in a spatially inhomogeneous medium in order to interpret the 
broad-band nonthermal radiation of the 
young shell-type supernova remnant (SNR) Cassiopeia A. A two-zone model is 
proposed that distinguishes between compact, bright steep-spectrum 
radio knots and the bright fragmented  
radio ring on the one hand, and the remainder of the shell - the 
diffuse `plateau' - on the other hand. In the framework 
of this model it is possible to explain the basic features of the spectral 
and temporal evolution of the synchrotron radiation of Cas~A if one assumes 
that these compact structures correspond to sites of efficient electron
acceleration producing  hard spectra of accelerated particles with 
power-law indices $\beta_{\rm acc} \sim 2.2$. The resulting 
energy distribution 
of radio electrons in these compact structures becomes significantly steeper
than the electron production spectrum on timescales of the energy dependent
escape of these electrons into the surrounding diffuse plateau region. We argue 
that  the steepness, rather than the hardness, of the radio 
spectra of compact bright structures in clumpy sources can in general 
be considered 
as a typical signature of sites where strong electron acceleration 
has built up high gradients in the spatial distribution of radio electrons.
Subsequent diffusive escape then modifies their energy distribution, leading 
to potentially observable spatial variations of spectral indices within 
the radio source. 
Qualitative and quantitative interpretations of a number of observational 
data of Cas~A are given. Predictions following from the model are discussed.
\end{abstract}

\section{Introduction}

Cassiopeia~A is the youngest of the known galactic supernova remnants (SNRs) 
whose birth probably dates back to 1680 (Ashworth \cite{ashworth}). 
It is one of the 
most prominent 
and well studied radio sources on the sky (e.g. Bell et al. \cite{bell75}; 
Tuffs \cite{tuffs86}; 
Braun et al. \cite{braun}; Anderson et al. \cite{arlpb} -- 
hereafter ARLPB; Kassim et al. 
\cite{kassim}, 
hereafter KPDE; etc.), whose synchrotron radiation  
probably extends into the hard X-ray region (Allen et al. \cite{allen}; 
Favata et al. \cite{favata}).  
Most of the radiation of both nonthermal and thermal origin 
comes from a shell region enclosed between two spheres, 
with angular radii $\simeq 150\,\rm arcsec$ and $\simeq 100\,\rm arcsec$, 
corresponding to spatial radii $R_0 = 2.5
\,\rm pc$ and $R_{\rm ring} = 1.7\,\rm pc$, respectively, 
for a distance of $d\approx 3.4\,
\rm kpc$ (Reed et al. \cite{reed}).
The former corresponds
to the mean radius of the assumed blast wave, while the 
latter is supposedly the mean radius of the reverse shock in the 
freely expanding (upstream) ejecta which are heating 
the gas to temperatures $\sim \! (1-3)\,\rm keV$, thus creating the hot 
thermal 
X-ray component (Becker et al. \cite{becker}; Fabian et al. \cite{fabian}; 
Jansen et al. \cite{jansen}).

Cas~A is considered to be a powerful particle accelerator, with 
an energy content in relativistic electrons, estimated from simple 
equipartition arguments, in the range 
$W_{\rm e}\sim 10^{48}-5\times 10^{49}\,\rm erg$   
(e.g. Chevalier et al. \cite{chevetal}; ARLPB ). This value should be increased 
by more than an order of magnitude for the total energetics in 
relativistic particles 
if the high ratio between relativistic protons and electrons observed in cosmic
rays (CR) holds in Cas~A. However, the mechanisms and sites of 
particle acceleration in Cas~A have not yet been identified.

Observations of Cas~A at  optical 
wavelengths show strong line emission from 2 main types of compact structures.
The quasi stationary flocculi are thought to trace the circumstellar medium
(e.g. Fesen et al. \cite{fesen}), while 
 numerous fast moving knots (FMK) represent  
dense clumps of supernova ejecta moving ballistically with average
velocities of $\simeq 5300\,\rm km/s $ from the time of explosion 
(see e.g. Reed et al. \cite{reed}, and 
references therein). The FMKs have been invoked as sites of particle 
acceleration by Scott \& Chevalier (\cite{scott75})
 who proposed second order Fermi acceleration of 
particles in the turbulence created in the wake of FMKs  
overtaking the shell. Another possibility is a first 
order Fermi acceleration process at the bow shocks driven ahead of FMKs  
as proposed by Jones et al. (\cite{jones94}). 
 These authors showed that these dense
gas `bullets'  can  effectively accelerate electrons at the stage of 
deceleration and subsequent fast destruction of the FMKs by Kelvin-Helmholtz 
and Rayleigh-Taylor instabilities.

\begin{figure*}
\resizebox{18.cm}{!}{\includegraphics{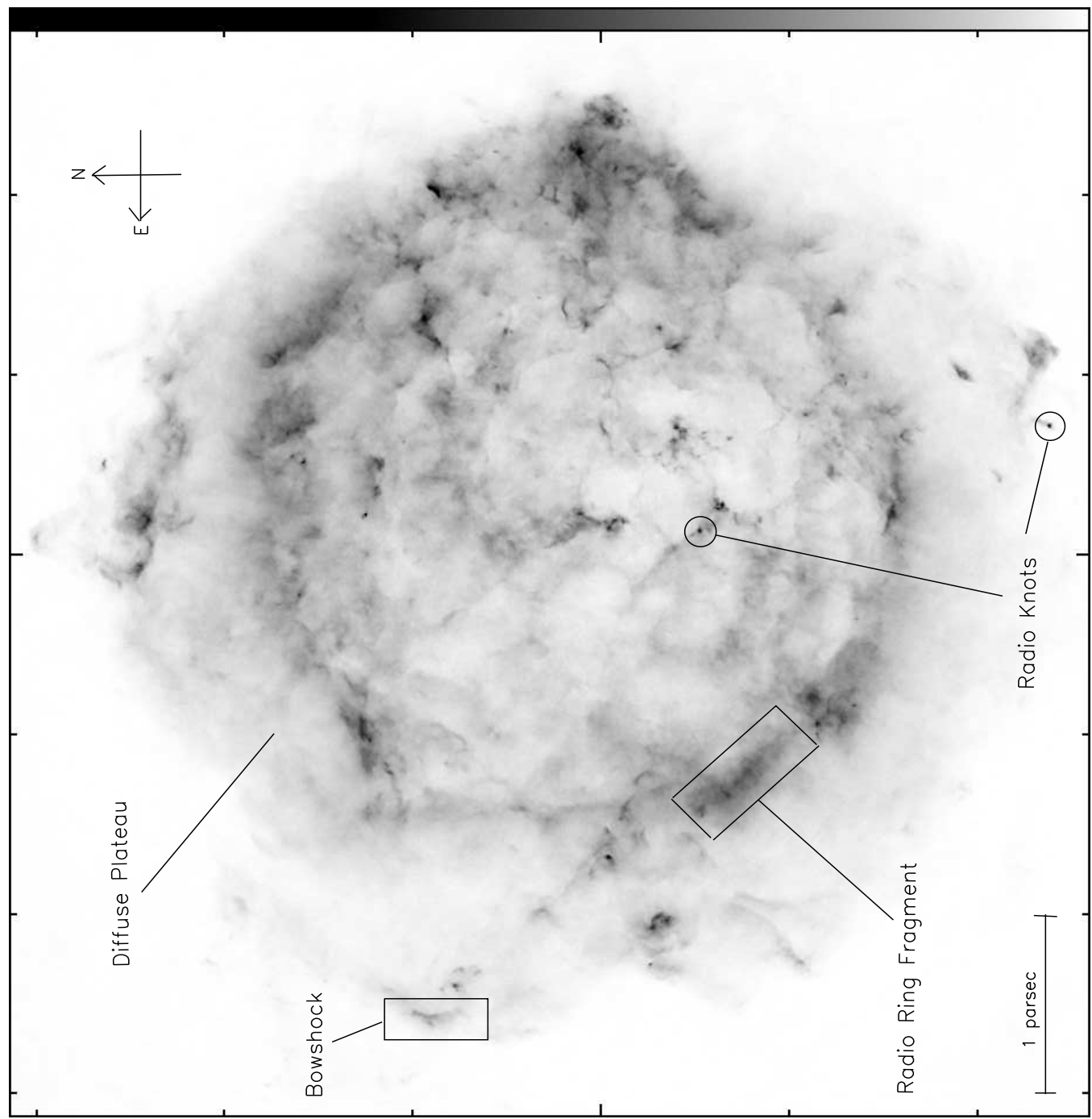}}
\caption{
VLA image of Cas A at $\lambda\,$6.3$\,$cm, showing examples of morphological 
classes of radio features referred to in the text. Axes (tangential projection
in RA, DEC for epoch B1950) are annotated at intervals of 1 pc (60.7 arcsec 
for the distance of 3.4$\,$kpc derived by Reed et al. \cite{reed}), 
with the central 
(longer) tick indicating the B1950 centre of expansion of the system of fast 
optical filaments at RA 23:21:12.0 DEC +58:32:17.9 (Kamper \& van den Bergh 
\cite{kamper}). The greyscale spans the full dynamic range of 0.0 to 12.09 mJy per 
0.375 arcsec pixel in 255 levels; a mild logarithmic scaling (0.85 exponent) 
has been applied to emphasise faint structures. The image was constructed 
from an almost fully sampled synthetic aperture using data from the VLA D, 
B, and A configurations at epochs 1983.0, 1984.0 and 1983.8, respectively 
(see table 1 of Braun et al. (\cite{braun}) for observational parameters). The 
brightness distribution has been deconvolved from the synthesised point 
spread function using the maximum entropy method (AIPS task VM).
 }
\end{figure*}

Perhaps the most straightforward suggestion, however, is that
efficient electron acceleration occurs directly in regions of high radio 
brightness -- in other words, that these regions are bright not only because
of enhanced magnetic 
field, but also due to local enhancement of relativistic electrons.
Such regions have been suggested to be  sites of electron 
acceleration by a number of authors (e.g. Bell \cite{bell77},  Dickel \& Greisen 
\cite{dickel}, 
Cowsik \& Sarkar \cite{cowsar84}).  
In Cas~A they are represented by rather compact 
structures which include  
$\geq 300$  radio knots (Tuffs \cite{tuffs86}, ARLPB) located in the 
shell, several paraboloidal radio features 
suggested to be bow shocks associated with decelerated ejecta 
(Braun et al. \cite{braun}), and the bright fragmented 
radio ring at the projected radius of the (supposedly) reverse shock/contact 
discontinuity. Examples of all these morphological structures are depicted 
in Fig.1.

Any theory of particle acceleration in Cas~A has to address a wide range of
spatial and spectral characteristics of the observed nonthermal
radiation from  the radio to infrared (IR), and possibly to X-ray regimes, 
which can be summarized as follows:

\begin{itemize}

\item For $20\,{\rm MHz} \leq \nu \leq 30\,\rm
GHz $ the total radio flux is well approximated by a power-law
$J(\nu) \propto \nu^{-\alpha }$ with an index $\alpha \approx 0.77$
(e.g. Baars et al. \cite{baars});

\item For $\nu \leq 20\,\rm MHz$ the spectrum $J(\nu)$ essentially flattens, and
  turns over below 15\,MHz (Baars et al. \cite{baars});

\item The secular decline of the fluxes   
 seems to be frequency-dependent, with a decline rate at the level
of $(0.8-1)\,\%/\rm yr$ in the frequency range 
$\nu \sim 40\,\rm MHz - 1\, GHz$
(Rees \cite{rees}; Hook et al. 1992), and $\sim 0.6\,\rm \%/yr$ at 
$\nu \sim 10\,\rm GHz$ (Dent et al. 
\cite{dent}; O'Sullivan \& Green \cite{osullivan}).

\item The contribution of the 
diffuse emission of the shell ({\it plateau}) to the total flux 
of Cas~A at 5\,GHz amounts to $\approx 50\,\%$, and  the second half is 
due to the fragments of  bright radio ring ($30\,\%$) and radio 
knots + bow shocks ($20\,\%$) (Tuffs \cite{tuffs86}). 
The total flux  at the epoch 1987 was about 750\,Jy (see ARLPB).

\item Individual radio knots show a wide spread of spectral indices in the 
range from $\alpha \sim 0.6 $ to $\alpha\sim (0.9-0.95) $ 
(Rosenberg \cite{rosenberg}; Tuffs \cite{tuffs83}; ARLPB; AR96).  

\item For $\nu \geq 30\,\rm GHz$ the radio spectrum flattens to $\alpha \approx 
0.65$ (Mezger et al. \cite{mezger}) which possibly extends to the IR region  
if the flux of continuum emission measured at 6\,$\mu \rm m$ by 
Tuffs et al. (\cite{tuffs97}) has a synchrotron origin.

\item  Recent observations reveal hard X-ray emission that extends 
with a power-law photon index $\alpha_{\rm x}\approx -3$  up to 120\,keV  
(Allen et al. \cite{allen}; Favata et al. \cite{favata}); 
a synchrotron origin of this
radiation implies acceleration of electrons 
to multi-TeV energies.

\end{itemize}

\noindent
Detailed studies based on high
resolution mapping of Cas~A at GHz frequencies  have suggested that the  
identification of bright compact radio features with electron 
acceleration sites can be problematic (ARLPB; Anderson \& Rudnick \cite{ar96}, 
hereafter AR96). 
A particularly interesting finding of ARLPB and AR96 is 
a significant and rather unexpected correlation between the spectral 
index of the knots and their projected  position in the shell, as well as
their  radio brightness: the steeper radio knots reside 
mostly in the outer regions of the shell, at $r\sim 2.5\,\rm pc$, 
 and tend to be brighter. 

The analysis in AR96 has shown that a self-consistent explanation of
 these correlations would be problematic if one assumes that the radio knots
 are the sites of efficient acceleration of the electrons.
For an SNR as young as Cas~A  the radiative energy losses 
cannot modify the spectrum of radio electrons. Therefore in the 
framework of a `standard' spatially homogeneous source model approach,
one has to 
attribute the radio spectral indices observed to the source spectra
of the electrons. 
The observed brightness/steepness trend of the radio knots would then 
apparently  rule out effective electron acceleration in 
the knots, because such a process implies a hardening 
(see e.g. Berezhko \& V\"olk \cite{berezhko}), rather than a
steepening of the particle spectra, typically to the 
power-law index $\beta \sim 2$. 

There does exist however a natural way to modify the 
spectra of radio electrons, if 
we abandon the standard approach of a spatially uniform 
source for treatment of these electrons.  
Spectral modifications become unavoidable if we take into account 
{\it energy-dependent} propagation and escape of relativistic particles
from the regions of higher concentration in an inhomogeneous medium.
The efficiency of this process depends on the spatial 
gradients in the energy distribution $N({\bf r}, E,t)$ of particles, 
and  timescales of the spectral modifications 
can be as short as the escape time $\tau(E)$. 
This effect is widely used for the interpretation of the galactic CR 
spectra in the framework of diffusive or Leaky Box models, 
but it has not yet been given proper attention in studies of radio 
emission in the CR sources themselves. 

In this paper we consider the consequensies of energy dependent 
propagation of relativistic electrons for the 
interpretation of the observed spectral, spatial and temporal 
characteristics of the broad-band nonthermal emission of Cas~A.
 In Sect. 2 we introduce the `two-zone' model for a 
 spatially non-uniform radio source, separating compact 
regions with a high density of relativistic electrons (zone 1) 
from the rest of the shell where the electron density is significantly lower 
(zone 2). In Sect. 2 we consider in a qualitative way possible 
consequensies of such 
an approach for the interpretation of the observed radio data. In Sect. 3 
we derive the system of kinetic equations for the electron energy distributions 
in the two zones. In Sect. 4 we 
assume that  zone 1 components correspond to the sites
of efficient electron acceleration, and show that
this scenario is able to explain the broad band non-thermal 
radiation data of Cas~A. In Sect. 5 we study the opposite scenario,
which assumes that the enhancement of the electron density in zone 1 is 
caused not by active acceleration of particles there, but rather only by 
compression of the background population of relativistic electrons in the shell. 
 We show that interpretation of the data within this latter 
scenario is problematic. 
In Sect. 6 we summarize the observational features which the model 
can explain, and in Sect. 7 we discuss implications and predictions of the 
model.

\section{The model }

\subsection{Insufficiency of the single-zone approximation}

To lowest approximation Cas~A might be represented as a 
homogeneous shell  
containing magnetic field, relativistic electrons and
gas. Although not reflecting the real pattern of the source, this 
`single-zone' approximation
allows  estimates of basic parameters such as 
 the mean magnetic field $B_0$ and energetics in radio electrons.  
In this approximation Cowsik \& Sarkar (\cite{cowsar80}) have derived a lower
limit to $B_0\geq 8\times 10^{-5}\,\rm G$,  comparing the 
expected bremsstrahlung flux of GeV electrons with the upper limit 
$I(> 100\,\rm MeV)\leq 1.1\times 10^{-6}\,\rm ph/cm^2 s$ of 
 SAS-2 (Fichtel et al. \cite{fichtel}) and COS~B detectors.

With the recent upper limits of the EGRET telescope,
$I(> \! 100\,\rm MeV)\leq 1.2\times 10^{-7}\,\rm ph/cm^2 s$ (Esposito et al. 
\cite{esposito}), this constraint
on $B_0$ is significantly strengthened. In Fig.\,2 we present the fluxes 
of the bremsstrahlung and synchrotron radiations 
calculated for 3 values of the mean magnetic field: 
$B_0=10^{-4}\,\rm G$, $\,3.5\times 10^{-4}\,\rm G$ 
 and $7\times 10^{-4}\,\rm G$. 
For the mean gas density (in terms of `H-atoms') we take $n_{\rm H}
=15\,\rm cm^{-3}$,  corresponding to $M \simeq 15\,M_{\odot}$ of
 matter in the shell (Fabian et al. \cite{fabian}, Jansen et al. \cite{jansen}, 
Reed et al. \cite{reed}),
and use a  mean atomic $\overline{Z (Z+1)/A} = 4.3$ 
derived by Cowsik \& Sarkar (\cite{cowsar80}) 
from the elemental abundance estimates 
of Chevalier \& Kirshner (\cite{chevkirsh}) in Cas~A. Continuous injection 
of electrons into the shell, with a spectrum 
$Q(E)\propto E^{-\beta_0} \,\exp (-E/E_{\rm c})$    
where $\beta_0 = 2\,\alpha_0 +1= 2.54$ and an exponential
cutoff energy $E_{\rm c}=100\,\rm TeV$, is assumed. 
The chosen $E_{\rm c}$ explains the fluxes 
of hard X-rays above 10\,keV as synchrotron emission 
for the case of $B_0 =10^{-4}
\,\rm G$. However, this value of $B_0$ is unacceptable 
because the bremsstrahlung flux produced by GeV radio electrons is too high. 
The EGRET upper limit for the $\gamma$-ray  flux of Cas~A requires
at least $B_0 = 3.5\times 10^{-4}\,\rm G$. Simultaneously we have to 
assume a significant flattening in the source spectrum 
of electrons below 100\,MeV in order to avoid contradictions with 
the observed X-ray fluxes at 100\,keV. 
For a single power-law distribution of electrons 
extending with $\beta_0 =2.54$ down to the MeV region, the 
lower limit to the magnetic field is $B_0=7\times 10^{-4}\,\rm G$.
Note however, that for $B_0 \geq 3.5\times 10^{-4}\,\rm G$
the radiative energy losses of the electrons establish a 
 spectral break at optical or lower frequencies, in which case 
the spatially homogeneous single-zone model fails to explain the 
observed hard X-ray fluxes by synchrotron radiation. 

\begin{figure}[htbp]
\resizebox{8.8cm}{!}{\includegraphics{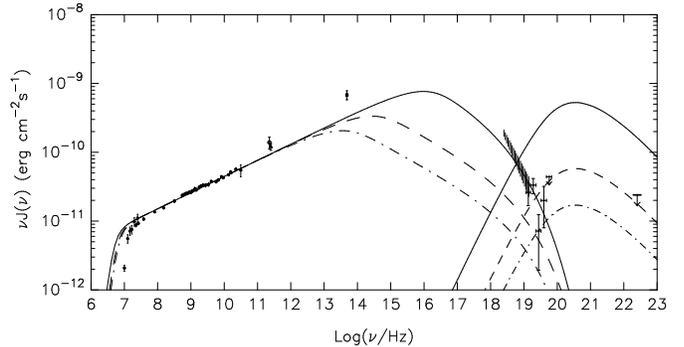}}
\caption{The fluxes of the synchrotron (heavy lines) and bremsstrahlung 
(thin lines) photons calculated in the framework of a spatially homogeneous
single-zone model for the source function of accelerated electrons with 
power law index $\beta_0=2.54$ for 3 different values of the magnetic field:
$B_0=10^{-4}\,\rm G$ (solid), $3.5\times 10^{-4}\,\rm G$ (dashed), and
$7\times 10^{-4}\,\rm G$ (dot-dashed). The radio flux 
measurements shown correspond to the data from Baars et al. (\cite{baars}) corrected
for the secular decrease to the epoch 1986. The data at 1.3\,mm (circle)
and 1.2\,mm (triangle) are from Chini et al. (\cite{chini}) and Mezger et al. 
(\cite{mezger}),
respectively. The hatched region corresponds to the range of X-ray fluxes 
measured by Allen et al. (\cite{allen}). Note that the RXTE fluxes are shown only 
above 10\,keV  since below 10\,keV the X-ray fluxes are dominated
by thermal emission. The OSSE measurements (The et al. \cite{the}) are shown by 
crosses, and the bar corresponds to the EGRET  upper limit above 
100\,MeV (Esposito et al. \cite{esposito}). 
 }
\end{figure}

\subsection{The two-zone model} 

A more realistic model for Cas~A has   
to account for the inhomogeneities in the radio brightness distribution
and for the 
observed spread in the power-law exponents of the radio
spectra of individual structures within  $\alpha \simeq (0.6-0.9)$.

The simplest approximation for a non-uniform radio source corresponds to a  
`two-zone' model, where the source with total volume $V_0$ consists
of 2 different regions, or zones, with volumes $V_1$ and $V_2=V_0-V_1$,
with 2 different spatial densities (concentrations) of 
radio electrons $n_1(E)=N_1(E)/V_1$ and $n_2(E)=N_2(E)/V_2$.
 Propagation effects may result in a significant difference between 
the two-zone and single-zone models only if these densities 
are essentially different. Therefore the principal definition for these two zones 
is that the concentration of electrons in 
zone 1 is much higher than in zone 2, $n_1\gg n_2$.
The energy distributions of the electrons in these zones 
can be different. Therefore in the power-law approximation 
$N_{1,2}(E)\propto E^{-\beta_{1,2}}$ the exponents $\beta_{1} $ and 
$\beta_{2} $ can be different.      
Because the spectral ratio $n_1(E)/n_2(E)$ may thus depend on energy,
it is convenient to specify that the condition $n_1\gg n_2$ 
relates first of all to electrons with energies $E\sim 1\,\rm GeV$ which 
are typically responsible for radio emission in the `1 GHz' band.

In order to apply the two-zone model to Cas~A, we have to 
foresee some observational manifestations of the radio emission of 
zones 1 and 2. It is reasonable to expect that not only the electron 
densities, but also the mean magnetic fields $B_1$ and $B_2$ in the two
zones differ, 
with the magnetic field presumably being higher in the regions of
higher electron densities. Therefore, the radio emissivities in these zones
will be strongly contrasting, with regions belonging to zone 1   
characterized by a much larger emissivity than regions in zone 2. It is also 
plausible that zone 1 components should be 
rather compact, with a small volume filling factor in the source, or otherwise
the overall emission would be strongly dominated by this single zone.

Application of this criterion to Cas~A immediately places  
all bright fragments of the radio ring in zone 1. 
Other constituents of zone 1 are the radio knots (and bow shocks) most of 
which are characterized by a strongly enhanced
emissivity.
We note however that in reality the knots embrace a wide 
range of emissivities, and because the simplified two-zone 
approach allows only 2 different electron densities (and emissivities),
some compact structures with relatively low brightness should be more 
correctly classified as belonging to zone 2. This may have a physical basis -- 
enhanced emissivity due to mild compression of magnetic fields, but without 
strong excess in the electon densities.
Zone 2 then consists of the rest of the (spherical) shell between the radio 
ring and the outer edge of Cas~A, i.e. the main body of the radio plateau, 
but also includes some low-brightness radio knots.

For purposes of practical classification of low-intensity knots
it may prove useful to resort also to the second potentially observable 
distinction between zones 1 and 2 -- different mean spectral indices
$\alpha_1$ and $\alpha_2$ of their intrinsic emission.  
Due to a faster escape of higher energy electrons from the regions of 
high concentration, one could generally expect that the radiation spectra 
produced in zone 1 should be  {\it steeper}, $\alpha_1 > \alpha_2$ 
(see below Sect. 2.2.3). Because
the spectral index of the total emission $\alpha_0\approx 0.77$ is to be
maintained, an immediate conclusion is that the radiation spectrum of
the diffuse zone 2 should be noticeably {\it flatter}, than the mean,
i.e. $\alpha_2 < \alpha_0  < \alpha_1$.  

 Because of the complicated morphology, 
a quantitative treatment of the spectral
indices of different types of structures is problematic in practice. 
In order to assess the characteristic values for $\alpha_{1,2}$,
we note that the line of sight integrated radio brightness seen 
towards individual radio knots shows a broad range of spectral
indices from $\alpha_{\rm min}\simeq 0.6$ to 
$\alpha_{\rm max}\simeq 0.95$, as measured 
by AR96 between $1.4\,\rm GHz$ and $5\,\rm GHz$.
These spectral indices however  do not always 
correspond directly to the {\it intrinsic} spectral indices of the knots,
due to background contamination from the plateau. Although 
the detailed least square fitting procedure for spectral index measurements used in 
AR96 (their Sect. 2.4) largely removes this problem for knots with high 
contrast to the plateau, for faint knots at the limit of confusion with
the plateau, the spectral index will be biased
towards  the (presumably flatter) index of that emission.  
In this regard, the result of AR96 presented in 
their Fig.6d appears as very informative. It shows that most of the radio knots 
found at large angular distances $\theta > 130^{\prime \prime}$
(the {\it projected} 
$r_{\perp}=2.17\,\rm pc$) 
belong to the steep-spectrum population: from 62 such knots in
total, 45 have indices $\alpha>0.8$, and  22 (i.e. $35\,\%$ !) belong to 
the `extreme' end of the observed spectral indices $0.86<\alpha<0.95$. 
Because at these
projected distances the brightness of the plateau is generally 
reduced\footnote{We  note that the low frequency radio maps of KPDE, with
angular resolution too poor to see individual radio knots, show a profound 
gradient of steepening spectral index of the plateau towards $\alpha \leq 0.9$
with projected radius $r>R_{\rm ring}$. It is plausible 
(though not quantitatively investigated) that this trend is due to
the population of steep spectra radio knots detected by AR96, which at lower 
angular resolution would contribute more flux than the plateau 
at angular distances $>130^{\prime \prime}$; the intrinsic spectrum
of the plateau emission might be then flat.}, this may 
indicate that the intrinsic spectral 
index of the radio knots on average could be much steeper than the 
 mean $\alpha_0$, and that the spectral index $\alpha > 0.86$
of the `extremely steep' population of the observed knots might be 
not so extreme, but rather a representative value for the intrinsic index 
of the knots. An (indirect) indication of this
suggestion could be also another result shown in Fig.~6d of AR96: 83 from 123 
knots found at angular distances close to the radio ring, 
$80^{\prime \prime} < \theta < 120^{\prime \prime}$, where the brightness of 
the plateau is relatively high, belong to the population of knots with
an `intermediate' steepness $0.71\leq \alpha_{\rm obs} \leq 0.8$. 
As discussed below in Sect. 6, the model does not exclude
a physical reason for such geometrical correlation between the knot index and 
its proximity to the ring. However, this effect can be explained also as 
a result of a line-of-sight mixture of {\it comparable} fluxes of the 
steep-spectrum knots with intrinsic $\alpha\sim (\alpha_0+0.1)$ 
and of the flat-spectrum diffuse plateau emission with $\alpha\sim (0.6-0.65)$
(see this Sect. below).
Moreover, if we believe that the real geometry
of the shell is quasi-spherical, it is difficult to suggest
any other reasonable explanation, except for invoking the flux contamination
effect,  why only 2 knots from those 123 show the index $\alpha_{\rm obs}>0.83$,
whereas the expected number of `extremely steep' 
knots to be located at deprojected radii $r>2.17\,\rm pc$ but at the angular
distances close to the ``ring'' should have been comparable with
\footnote{Even if the steep spectrum knots were physically distributed on the
outermost radius of Cas~A ($\simeq 150^{\prime \prime}$), we would still expect
significantly more than 2 knots to be seen within projected distances
$80^{\prime \prime} < \theta < 120^{\prime \prime}$}  the number $22$.
    
We can conventionally define that all knots which have a moderate brightness
and show spectral index $\alpha \leq 0.7$ 
belong to zone 2.  Then, given the tendency for brighter knots to be steeper
(AR96), this classification would retain most of the radio knots (at least in 
terms of their overal flux) in zone 1. At the same time, excluding thus the 
flat-spectrum knots from zone 1, for the average intrinsic spectral index of
the remaining population of knots one can reasonably suppose  
$\alpha_{1,\rm kn}\sim (\alpha_0+0.1) = 0.87$ or so.  
  
It is possible that the intrinsic spectral index of the zone 1 components
identified with the bright fragments of the radio ring is also close 
to this value. The low frequency measurements of     
Woan \& Duffet-Smith (\cite{woan}; hereafter WDS90) at angular resolution 
$\geq 14^{\prime \prime}$ show that the average spectral index of 
the bright large radio components correlated with the ring  
is $ \simeq 0.83$, i.e. again significantly larger than the 
mean $\alpha_0$ of the total emission. If we accept that the contribution
of the diffuse emission, zone 2, may be flat, the intrinsic spectral index 
of the fragmented radio ring should be even larger.   
Then  $\alpha_1\sim (\alpha_0+0.1)$ may be a reasonable 
value for the average spectral index of the entire zone 1. For qualitative 
discussions in this section we will use $\alpha_1=0.87$. We note however that
our model does not pretend to predict characteristic
spectral indices with the accuracy of the second digit after the comma. 

The first zone will thus include a number $K \sim 200$ individual 
components, i.e. basically the bright fragmented radio ring, 
the `radio' bow shocks, and the main part of all radio knots. 
Besides, zone 1 could also contain some regions of high concentration 
of electrons and high emissivities which are not immediately distinguished 
on the intensity maps  either because of their very small sizes or 
because of strong 
contamination by, and confusion with the diffuse plateau emission. The latter 
may be the case if we believe that the fragmented ``radio ring'' actually 
represents a chain of large but geometrically flat (thin) structures of the 
``radio sphere'' which are observed mostly `edge-on'. Then at angular distances 
interior to the ring, $\theta < 80^{\prime \prime}$ or so, such fragments  
observed `face-on' may easily merge with the plateau emission and show up as 
diffuse `clouds' of only moderately enhanced brightness (see Fig. 1).  

The flux produced in zone 1 can be estimated if we 
remember that at 5 GHz the ({\it background subtracted}) flux of the radio 
ring fragments makes up $\approx 30\%$, and that 
radio knots produce $\approx 20\%$  
of the total flux (Tuffs \cite{tuffs86}). The latter value should
be somewhat reduced for the flux of knots to be removed from zone 1 and 
classified in zone 2. On the other hand, it seems quite possible
 that up to $(5-10)\,\%$ of the total flux should be assigned to the `unseen' 
population of zone 1 (especially, to face-on structures of the 
``radio sphere''). Therefore, we can estimate the overall flux $J_1$ of zone 1 as 
$\sim (45-55)\,\%$ of the total, concluding that the fluxes of 
zone 1 and zone 2 at 5 GHz are about the same, $J_1\simeq J_2$.

The spectral index $\alpha_2$ in the diffuse zone 2 can be then predicted
if we note that the total flux has the form
\begin{equation}
J(\nu)= J_1 \left( \frac{\nu}{5\,{\rm GHz}}\right)^{-\alpha_{1}} +
J_2 \left( \frac{\nu}{5\,{\rm GHz}}\right)^{-\alpha_{2}}\; .
\end{equation}
In order for the mean spectral index 
between $\nu_{\ast}=1.4\,\rm GHz$ and $\nu_{\ast \ast}=5 \,\rm GHz$  
to be equal to $\alpha_{0}$,  one needs a flux ratio  
\begin{equation}
\frac{J_2}{J_1} = \frac{(\nu_{\ast \ast}/\nu_{\ast})^{\alpha_{1} - \alpha_0}- 1}
{1 - (\nu_{\ast\ast}/\nu_{\ast})^{\alpha_{2}-\alpha_0}}\;\; \cdot
\end{equation}
Then for $J_2 = J_1$, $\alpha_0=0.77$, and $\alpha_{1} =0.87$,    
we find $\alpha_{2}= 0.65$. It may somewhat vary assuming 
slightly different $J_2/J_1$ or $\alpha_1$ (or $\alpha_0$, which can be
also slightly less than 0.77, see Fig.2). 
In particular, $\alpha_{2}$ 
may be rather close to 0.6 if we allow for the idea that the `unseen'
face-on structures
of the ``radio sphere'' could contribute into zone 
1 a third of what is contributed by the edge-on ``radio ring''.  
Remarkably, these values of $\alpha_{2}$ correlate well with the lower 
range of spectral indices of the flat-spectrum radio knots.
This suggests that flat-spectrum radio knots, with spectral indices
close to 0.6 (or flatter than 0.7 as classified above) may indeed  
represent the regions of moderate local compression
of the magnetic field and/or relativistic electrons of the diffuse shell, 
and therefore may give information about the spectrum of
the background population of electrons there.

Thus, the  model suggests a rather flat intrinsic spectrum of the radio plateau
in the range $\alpha_2\simeq 0.6-0.65$, which corresponds to the spectral index of 
electrons $\beta_2\simeq 2.2-2.3$. For zone 1, $\alpha_1\simeq 0.87$ implies 
a rather steep electron spectrum with $\beta_1\simeq 2.74 $.
Taking into account the enhanced brightness of the compact zone 1, 
we will refer to zone 1 structures  as the compact bright steep-spectrum 
radio (CBSR) components.

\subsubsection{Synchrotron self-absorption in the two-zone model}

Even though still highly simplified, this two-zone approach suggests a 
qualitative interpretation of several observed radio features of 
Cas~A. In particular, it allows us to solve  the problem 
of the low-frequency turnover below 20\,MHz. 

Although the measurements of KPDE at $\sim 100\,\rm MHz$
show evidence for thermally absorbing low-temperature gas 
in the central $\leq 1\,\rm pc$ region of Cas~A, 
the thermal absorption cannot result in any noticeable effect 
in the shell where 
$T\sim 10^7\,\rm K$. On the other hand,
in the case of a spatially homogeneous source (the shell) with a mean field 
$B_0 = 0.7\,\rm mG$, the process of synchrotron self-absorption results in a 
turnover of the radio spectrum only at frequencies below 5\,MHz (see Fig.2).
To explain the observed turnover below $20\,\rm MHz$,
 in a single-zone approach one needs  $B_0\geq 5\,\rm mG$. 
But this assumption boosts the magnetic 
field energy up to $W_{\rm B}\geq 10^{51}\,\rm erg$, and therefore 
is hardly acceptable. 
 
The situation essentially changes if one takes 
spatial inhomogeneities in the distribution of radio electrons and magnetic 
fields into account. Indeed, in the framework of the proposed  
two-zone approach the radio flux at frequencies well below 1 GHz  
is dominated by the steep-spectrum component, i.e. one has to
require synchrotron self-absorption mainly of the zone 1 flux at 
$\nu < 20\,\rm MHz$.
Because of the very high density of the radio electrons and of the magnetic field 
in compact zone 1 components, synchrotron 
self-absorption there should be expected  
at frequencies significantly higher than in the single-zone model.

In order to estimate the model parameters needed for synchrotron self-absorption
to occur below $20\,\rm MHz$, we note that for a power-law distribution 
of electrons 
$N(E) \propto E^{-\beta}$  the synchrotron luminosity can be expressed as
\begin{eqnarray}
L(\nu) &= &2\times 10^{-25}\,C_{\beta}\, N_{\ast}\, 
\left( \frac{B}{1\,\rm mG} \right)^{\frac{1+\beta}{2}}\times \nonumber \\
& & \left( \frac{\nu}{10\,{\rm GHz}}\right)^{\frac{1-\beta}{2}}\; \;
\rm \frac{erg}{s\,Hz}\; ,
\end{eqnarray}
where $N_{\ast} \equiv \, E_{\ast}\,N(E_{\ast})$ with $E_\ast =1\,\rm GeV$.
$C_{\beta}$ is a weak function of $\beta$ that changes within $1\pm 0.15$
for $\beta \sim (2-3)$, and we will assume $C_{\beta}=1$.
With the same accuracy, the synchrotron absorption coefficient is given by 
\begin{eqnarray}
\kappa(\nu)& \simeq & 9 \times 10^{-17} \, \beta \,10^{3 \beta/2}\,
\frac{N_{\ast}}{V}\,\left(\frac{B}{1\,\rm mG}\right)^{1+\beta /2} 
\times \nonumber \\
& &\left( \frac{\nu}{10^7\,\rm Hz}\right)^{-(2+\beta /2)} \; \rm cm^{-1} 
\, ,
\end{eqnarray}
where $V$ is the volume in units of $\rm cm^3$.
The volume of zone 1 can be represented in the form 
$V_1 =\sum_{i=1}^{K} V_{1}^{(i)} \simeq  \overline{S}_{\rm pr} \bar{d}$, 
where $\bar{d}$ is the mean thickness  of individual CBSR components 
along the line of sight and $ \overline{S}_{\rm pr} = \sum_{i=1}^{K} 
\overline{s}_{i}$ is their overall surface in the plane of the sky.  
Requiring the synchrotron 
opacity to be significant at $\nu < 20\,\rm MHz$ (which implies 
$\kappa \, \bar{d} \simeq 1$  at $\nu \simeq 15 \,\rm MHz$) we obtain 
\begin{equation}
  \overline{S}_{\rm pr} \simeq 1.3\times 10^{37} 
\left( \frac{J_1}{400\,\rm Jy}\right)
\, \left(\frac{B_1}{1\,\rm mG}\right)^{0.5}\; \rm cm^2 \,.
\end{equation}
For the external radius of the plateau region $R_{\rm 0} =2.5\,\rm pc$ 
this leads to an area filling factor
\begin{equation}
\frac{\overline{S}_{\rm pr}}{S_{0}} \simeq 0.07 \, 
\left( \frac{J_1}{400\,\rm Jy}\right)
\, \left(\frac{B_1}{1\,\rm mG}\right)^{0.5}\; ,
\end{equation}
where $S_0=\pi R_{0}^2$.
For   $B_1 \simeq 
(1-2)\,\rm mG$ in zone 1 (see Sect. 4), 
the ratio  $\overline{S}_{\rm pr}/S_{0} \sim 0.1$. 
From Fig.1 it is apparent that
the area filling factor of all bright radio features is indeed of the order of 
10\,\%\,. 

From Eq.(5) it follows that the characteristic length scale for 
$K\sim 200$ CBSR components of the zone 1 
should be of order $\Delta l \sim (\overline{S}_{\rm pr}/K)^{1/2} 
\sim 0.1\,\rm pc$. 
If we separate the radio knots from the larger fragments of the 
radio ring, and take into account that the flux $J_{\rm kn}$
of the knots constitutes about a third of $J_1$, the  mean radius of 
the knots can be estimated as $\sim 0.03 \,\rm pc$. At 
the distance 3.4\,kpc  this corresponds to the  
angular sizes of about $2^{\prime \prime}$, 
very similar to what is really observed.

\subsubsection{Flattening of the radio spectrum} 

The superposition of flat and steep spectral 
components, with $J_1\simeq J_2$ around $5\,\rm GHz$, results in a flattening 
of the total radio spectrum at higher frequencies. 
Observations of Mezger et al. (\cite{mezger}) do indicate a  
noticeable flattening of the Cas~A spectrum in the wavelength 
range from $1\,\rm cm$ to 1mm, 
with power law index $\simeq \! 0.65$ that possibly extends up to 
the near-IR region      
if the flux of continuum emission measured at $6 \,\mu \rm m $ 
by Tuffs et al. (\cite{tuffs97}) 
is synchrotron in origin (see Fig.2).

Another important effect related to such composite fluxes consists in the
possibility to give a new interpretation of the claimed secular 
flattening of the radio spectra (Dent et al. \cite{dent}; O'Sullivan \& Green 
\cite{osullivan}). 
Namely, the flattening of the radio spectra
in time could be easily explained  
as being due to different rates of decline of the fluxes of the two zones,
with the plateau emission dropping more slowly. 
Observations at 5\,GHz have shown that the CBSR components are fading more 
rapidly than the total emission, which then requires that the
plateau must be fading less rapidly (see Table 6.1 of Tuffs \cite{tuffs83}). 
If the 
plateau indeed were to have a flatter spectral index than the CBSR components,
this would result in a flattening of the spectral index of Cas A with time.
Note that previously, in the framework of the single-zone
approach, the secular flattening could be explained only 
by the assumption of a gradual flattening of the spectrum of 
electrons either at higher energies or in time (Scott \& Chevalier \cite{scott75}, 
Chevalier et al. \cite{chevetal}, Cowsik \& Sarkar \cite{cowsar84}, 
Ellison \& Reynolds \cite{ellison}).

\subsubsection{Effects of energy dependent propagation}

Another advantage of a spatially 
inhomogeneous model is its ability to explain
the basic characteristics of the observed nonthermal radiation in terms 
of very efficient acceleration processes operating in Cas~A,   
which generally result in a source function  
of accelerated particles with  hard spectral indices 
$\beta_{\rm acc} \simeq (2-2.2)$ (e.g. efficient Fermi acceleration 
at strong shock fronts). Even though the spectral index 
formally corresponding to the mean $\alpha_0= 0.77$ is $\beta_0 \geq 2.5$,
the assumption of $\beta_{\rm acc}$ much harder than $\beta_0$ becomes 
possible if one takes into account the  energy dependent propagation 
and escape of electrons  from regions of high concentration
on timescales $\tau_{\rm esc}(E)\propto E^{-\delta}$. 
For example, assuming that 
the sites of efficient particle acceleration are located in 
zone 1, one could expect that 
the electron energy distribution $N_{1}(E)$ formed there on timescales 
$t\gg \tau_{\rm esc}$ would have the power-law index 
$\beta_{1}  = \beta_{\rm acc} + \delta$. On the other hand the leakage from  
zone 1  corresponds to injection of relativistic
electrons into surrounding zone 2
with the rate $Q_2(E)= N_{1}/\tau_{\rm esc} \propto E^{-\beta_{2} }$ where  
$\beta_{2}  = \beta_{1}  - \delta = \beta_{\rm acc}\,$.  
With a reasonable source function index
$\beta_{\rm acc} \simeq 2.2-2.3$ and energy-dependent escape with 
$\delta \sim 0.5$, the values of $\beta_{1}$ and $\beta_{2}$ 
needed for interpretation of the Cas~A 
radio data could be reproduced.

Thus,  the energy distribution of electrons $N_1(E)$
in the compact regions of efficient acceleration, which results in the 
enhanced density of relativistic electrons, 
may be actually much steeper than $N_2(E)$ in the surrounding volume.  
Interestingly, although the electron distribution $N_2(E)$ in the plateau 
region would show the hard spectral index of acceleration $\beta_{\rm acc}\,$,
in principle no acceleration of electrons needs occur there at all. 
It is 
important to note, however, that on a qualitative level 
the same power-law indices $\beta_{2} $ and $\beta_{1} $ for both zones could be 
expected also in the case when relativistic electrons were accelerated in 
zone 2, while the electron density in 
CBSR structures would be enhanced due to local compression of the electrons,
but not efficient acceleration in zone 1.
To distinguish between these 2 basic scenarios, 
quantitative modelling is needed.

\section{Kinetic equations in the two-zone model}

Approximating the  momentum distribution function of relativistic electrons 
as an isotropic function $f_{p}$ of the momentum 
$p= |\,{\bf p }\,|\,$, 
the kinetic equation for the energy distribution  $f \equiv 
f({\bf r},E,t){\rm d}E = 4\pi p^2 f_{p}{\rm d}p$ of relativistic particles 
($E\simeq p c$) can be written as:
\begin{eqnarray}
\frac{\partial f}{\partial t} & = & {\rm div}_{\bf r} 
(D\,{\rm grad_{\bf r}} f) 
- {\rm div}_{\bf r}({\bf u} f) + 
\frac{{\rm div}_{\bf r}{\bf u}}{3}\frac{\partial}{\partial E}(E f) 
\nonumber \\
& &
+ \frac{\partial}{\partial E}(P f) + A[f] \; .
\end{eqnarray}
Here  ${\bf u} \equiv {\bf u}({\bf r},t)$ is the fluid velocity, 
$P\equiv P({\bf r}, E, t)$ is the energy loss  rate of electrons.
Assuming isotropic diffusion, $D\equiv D({\bf r}, E, t)$  
is the scalar spatial  diffusion 
coefficient, whereas  
$A[f]$ is a functional standing for various stochastic acceleration terms of 
electrons.

In the two-zone model the source with volume $V_0$  
is subdivided into zone 1 composed of $K$ compact structures, with   
volume $V_1 =\sum_{i=1}^{K} V_{1}^{(i)} \ll V_0$, spatially 
immersed in the zone 2 with 
$V_2 =  V_0 - V_1 \approx V_0$. The set of equations describing the evolution 
of the total energy distribution function of particles in these two zones,
$N_{j}(E,t)={\int} f \,{\rm d} V_j\,$, can be found by integration of 
Eq.(7) over $V_j\,$ ($j=1,2$). 
Integration of the term $\partial f / \partial t$ over $V_1$ results in
$\partial N_1 / \partial t$. 
The volume integral of the two first terms in the right hand side of Eq.(7)  
is reduced to the sum of $K$ integrals on
the surface $S_{1\,2} = \sum_{i}^{K} S_{i}$ separating the zone 1 components
from zone 2:    
\begin{eqnarray}
\int_{ V_1} [ {\rm div}_{\bf r} (D \,
{\rm grad_{\bf r}} f)  -  {\rm div}_{\bf r}({\bf u} f) ]\,{\rm d}^3 r 
& = &  \\ 
 \sum_{i=1}^{K}\; [\, \oint_{S_i} 
D\,({\bf e}\, {\rm grad}_{\bf r} f)\, {\rm d} s 
& - & \oint_{S_i} ({\bf e \, u}) f  \, {\rm d} s \,] \; . \nonumber
\end{eqnarray}
Here {\bf e} is  the unit 
vector perpendicular to the 
surface element ${\rm d}s$,  directed outward from zone 1 to zone 2. 
These terms describe the rate of diffusive and convective exchange of 
particles between the two zones. 

To proceed further, let us separate those parts of the interface $S_{1\,2}$ 
where the scalar product ($\bf u \, e$) has positive and negative signs:
$S_{1\,2} = S_{+} +S_{-}$. The surface $S_{+}$ corresponds to the 
regions of plasma outflow from zone 1 into zone 2, so the  
integral over $S_{+}$ of the second term in Eq.(8) describes the    
convective escape of electrons, $- N_1(E,t)/\tau_{\rm c}$, 
where 
\begin{equation}
\tau_{\rm c} \simeq \bar{h}/{ u_1}\; ,
\end{equation} 
$\bar{h}$ is characteristic 
thickness of, and $u_1$ is the plasma velocity in CBSR components.

The first integral in Eq.(8) over the surface $S_{+}$  
can be simplified if we approximate    
$({\bf e} \,{\rm grad} f) \simeq (\bar{f}_2 - \bar{f}_1)/\Delta l\,$,
where  $\bar{f}_{1}\equiv \bar{f}_{1}(E,t)$ and $\bar{f}_{2}=
\bar{f}_{2}(E,t)$ 
are the volume averaged distribution functions of relativistic particles in 
zones 1 and 2, respectively, and $\Delta l\equiv \Delta l(E,t)\,$ is the 
characteristic  thickness of the transition layer between these zones.
Taking into account that the volume of a body can be expressed as
$V= a h S_{+}$, where $h$ is the thickness and 
$a\sim 1$ is a factor depending on the shape of the body, 
we find 
\begin{eqnarray}
\sum_{i=1}^{K}\oint_{S_{i+}} D\,({\bf e} \, {\rm grad} f )& = &
\sum_{i=1}^{K} \frac{\bar{f}_{2} -\bar{f}_{1}}{\Delta l_i }
\, \overline{D}_{ i} \, S_{i+}  \nonumber \\ & = &
\frac{\bar{f}_{2} - \bar{f}_{1}}{\tau_{\rm dif}} \, V_1 \; ,
\end{eqnarray}
where $\tau_{\rm dif} $ has the meaning of characteristic diffusive escape 
time of relativistic electrons  from the regions of higher  
electron densities:
\begin{equation}
\frac{1}{\tau_{\rm dif}(E,t)} = \sum_{i=1}^{K}
\frac{ \overline{D}_{ i }}{a {h}_{i} \Delta l_i} \,
\frac{V_{1}^{(i)}}{V_1}
\rightarrow \frac{\overline{D}_{\rm r}(E, t)}{a \bar{h} 
\Delta \bar{l}(E,t)}\; .
\end{equation}
Here $\overline{D}_{\rm r}(E, t)$ corresponds to the
mean of the spatial diffusion coefficient ${D}_{\rm r}({\bf r},E,t)$ 
on the surface $S_{+}$.
Taking further into account that 
$N_{j} = V_{j} \, \bar{f}_{j}$~ ($j=1,2$), the diffusive escape 
({\it exchange}) term is reduced to the form 
\begin{equation}
\frac{V_1 \, N_{2}(E,t)}{V_2 \, \tau_{\rm dif}(E,t)} - 
\frac{N_1(E,t)}{\tau_{\rm dif}(E,t)} \; \cdot
\end{equation}

\vspace{2mm}

The integrals over the surface $S_{-}$ in Eq.(8) describe the 
inflow of electrons from zone 2, with some rate
$Q_{2\, 1}$, through the frontal surfaces of zone 1 components. 
For a mainly adiabatic compression of
relativistic particles (see Sect. 5), $Q_{2\, 1}$ 
could be directly 
connected with the energy distribution $N_2$ of electrons in 
 zone 2. One could 
 however expect the formation of strong shocks due to fast motion
of CBSR components with respect to the surrounding plasma, so that 
the upstream electrons could be accelerated  
before entering zone 1. In that case $Q_{2\, 1}$ would correspond
to the source
function of electrons accelerated on the shock fronts and injected into 
zone 1. 

An essential part of the diffusive shock acceleration process is given by the
adiabatic compression term in Eq.(7) (third term) in the shock region near 
the surface $S_{-}\,$. Therefore the volume integral of that term in a 
thin region adjacent to $S_{-}\,$ is to be formally incorporated into the
source function $Q_{2\, 1}$. The integral in the main part of the volume 
$V_1$ describes adiabatic energy losses (or gain) of electrons,
and can be combined with the volume integral of the forth term in Eq.(7) as:
\begin{equation}
\frac{\partial}{\partial E} \int_{V_1} \left( E {\rm div}_{\bf r}{\bf u}/3 +
P \right)\,f \,{\rm d}^3 r  = 
\frac{\partial}{\partial E} \left( \,\overline{P}_1\,
N_1\right)\;,
\end{equation}
where $\overline{P}_1 \equiv \overline{P}_1(E,t)$ is the mean rate of 
energy losses of 
electrons, which now includes also the adiabatic energy loss term
$\overline{P}_{\rm ad}\simeq E\oint ({\bf u\,e})\,{\rm d}s/3$. 

Finally, the volume integral
of the last term in Eq.(7) describes internal sources of accelerated
electrons 
$Q_{1}^{\rm (int)}$ inside the 
volume $V_1$. The final equation then reads:
\begin{equation}
\frac{\partial N_1}{\partial t} = \frac{\partial (\,\overline{P}_1\,N_1\,)}
{\partial E} \, - \, \frac{N_1}{\tau_{\rm esc}}\,  
+\, \frac{V_1\,N_2}{V_2\,\tau_{\rm dif}}\, +\, Q_1 \; .
\end{equation}
Here $Q_1 \equiv Q_1(E,t) = Q_{2\,1}+ Q_{1}^{ \rm (int)}$ corresponds to 
the total injection rate of electrons in the first zone,  
and $\tau_{\rm esc}$ is the overall  
"diffusive + convective" escape time of particles 
\begin{equation}
 \tau_{\rm esc}(E,t) = \left[ \frac{1}{\tau_{\rm dif}(E,t)}
+\frac{1}{\tau_{\rm c}(t)}\right]^{-1}\; .
\end{equation}

Because $\tau_{\rm dif}$ generally increases 
for decreasing $E$,  $\tau_{\rm esc}$ becomes
energy {\it independent} at sufficiently small energies when
$\tau_{\rm dif} \geq \tau_{\rm c}$. The range of actual energy dependence of 
$\tau_{\rm esc}$ is limited also at high energies 
since the diffusive escape time 
cannot be less than the light travel time across the CBSR  
components $\sim \bar{h} / c$. 
This obvious requirement formally
follows from the condition that the characteristic 
lengthscale $\Delta \bar{l}(E)$ for spatial gradients in the 
distribution function $f({\bf r}, E, t)$ 
cannot be less than the mean electron scattering path 
$\lambda_{\rm sc}(E)$, since otherwise the diffusion approximation 
implied in Eq.(7) fails. 
Taking into account that $D \sim \lambda_{\rm sc} c /3$, from Eq.(11)
we find that indeed 
$\tau_{\rm dif}(E,t) \geq \tau_{\rm min} \simeq 3 a \bar{h}/c$.
Assuming 
$D(E)\propto E^{\, \delta_{1}}$,
a reasonable approximation for the diffusive escape time is then  
\begin{equation}
\tau_{\rm dif}(E,t) = \tau_{\ast}(t) \,(E/E_{\ast})^{- \delta}
\, +\, \tau_{\rm min}(t)\; .
\end{equation}

From Eq.(11) follows that generally the power-law index 
$\delta$ for diffusive escape time  should be smaller than the   
index $\delta_1$ of the diffusion coefficient: since faster 
diffusion of more energetic particles tends to 
smooth out the gradients of the distribution function 
$f({\bf r}, E, t)$ more effectively, the characteristic 
lengthscale $\Delta \bar{l}(E,t)$ should increase with energy. 
Then for a power-law approximation $\Delta \bar{l}
\propto E^{\,\delta_2}$ the index $\delta_2 >0$, 
so $\delta = \delta_1 - \delta_2 < \delta_1 $.   
This conclusion is important since it allows us to assume a power-law index
$\delta \sim 0.5-0.6$ needed for modification of the electron spectral index, 
not excluding at the same time efficient shock acceleration of 
particles in the Bohm diffusion limit which corresponds to $\delta_1 = 1$. 

Integration of Eq.(7) over the volume of zone 2   
results in an equation for $N_2(E, t)$:
\begin{equation}
\frac{\partial N_2}{\partial t} = \frac{\partial (\,\overline{P}_2\,N_2\,)}
{\partial E} \, + \, \frac{N_1}{\tau_{\rm esc}}\,  
-\, \frac{V_1\,N_2}{V_2\,\tau_{\rm dif}}\, +\, Q_2 \; ,
\end{equation}
where $\overline{P}_2$ is the mean energy loss rate, and 
$Q_2$ is the source function of electrons in zone 2.
The second and third terms here result from the same
surface integrals on the interface between zones 1 and 2 
as in Eq.(8), but
they have opposite signs since now the direction of the unit vector {\bf e}  
is reversed. Note that formally the volume integrals
in Eq.(8) result in two additional surface integrals
describing particle exchange
between zone 2 and the exterior on its 
 outer boundary. However, this actually 
extends the two-zone model to a three-zone one, therefore we neglect here 
these exchange
terms in order to remain in the framework of the two-zone approach.

The set of kinetic equations (14) and (17) describes the evolution of 
relativistic  electrons 
in the general case of an expanding inhomogeneous source when the electron 
energy losses and particle exchange rates between two zones are 
time-dependent. Given these parameters and the 
source   functions $Q_{1,2}(E,t)$, the electron energy distributions  
can be found numerically by iterations, 
using analytic solutions to the kinetic equation
of electrons in an expanding homogeneous medium (Atoyan \&
Aharonian \cite{aa99}). To reduce the number of free parameters, 
we shall consider basically the case of a  stationary source 
because the current energy distributions depend mostly on the injection
of the electrons during recent (largest) timescales,  
$\Delta t_0\simeq 300\,\rm yr$. Only for calculations of the secular decrease 
of the fluxes we shall use the general results for an expanding source, 
taking into account adiabatic losses and decreasing average magnetic fields.

\section{CBSR components as electron acceleration sites}

Let us first consider the hypothesis that the compact bright 
radio components are the main sites of electron acceleration, and 
neglect possible acceleration
in zone 2, i.e. $Q_2=0$. The electron injection function in zone 1 
is approximated as 
\begin{equation}
Q(E) = \, Q_0\, E^{-\beta_{\rm acc} }\exp(-E/
E_{\rm cut})\; ,
\end{equation}
where we have omitted the suffix `1' in $Q$.
For calculations we assume equal radii $R_{\rm 1}$ and magnetic fields 
$B_1$ for  zone 1 components.

Since the thickness  $\bar{h}$ of a spherical object
corresponds to its diameter, Eq.\,(9) gives 
 $\tau_{\rm c}=2 R_1/u_1$. For $R_1\sim (0.05-0.1)\,\rm pc$, and  
$u_1$ less than the sound speed
$v_{\rm s}= \sqrt{\gamma_{\rm adb} k T/A\, m_{\rm p}}
\sim 150 \,\rm km/s$ in the gas with $T_1\leq 3\times 10^7
\,\rm K$ and mean atomic $A\sim 15$ (corresponding to $\overline{Z (Z+1)/A} 
\simeq 4.3$, Cowsik \& Sarkar \cite{cowsar80}), 
the time $\tau_{\rm c}$ is larger than the 
source age $t_0\simeq 300\,\rm yr$. Therefore the precise value of 
$u_1$ will not significantly affect the results. 
The minimum escape time of electrons from the
zone 1 is taken as $\tau_{\rm min} =R_1/c$. 

%
\begin{figure}[htbp]
\hspace{10mm}\resizebox{6.5cm}{!}{\includegraphics{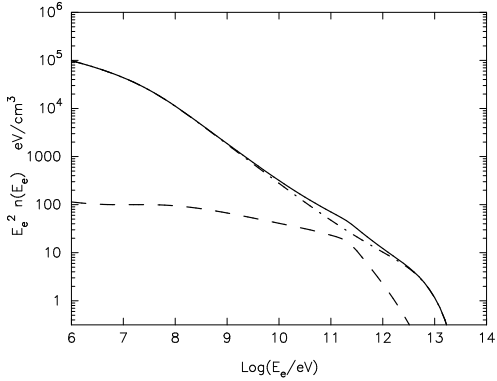}}
\caption{The energy distributions of the electrons in zone 1 (solid) and
zone 2 (dashed), calculated in the two-zone model
assuming that the CBSR components in CAS~A correspond to the
sites of electron acceleration. The dot-dashed curve corresponds to the 
initial iteration for the energy distribution of electrons in zone 1
when the `return' diffusive flux of electrons from zone 2 into zone 1 is
neglected (the term $\propto N_2/\tau_{\rm dif}$ in Eq.14).
Parameters for the source function 
are $\beta_{\rm acc}  =2.24$, $E_{\rm c}=3.5\times 10^{13}\,\rm eV$; 
the magnetic fields  
 $B_1=1.5\times 10^{-3}\,\rm G$ and 
$B_2=4\times 10^{-4}\,\rm G$; the diffusive escape time is defined by 
 $\tau_{\ast}=25\,\rm yr$ and $\delta =0.6$. For zone 1, 
$K=200$ components with the radius $R_1=0.05\,\rm pc$ 
are supposed.}
\end{figure} 

The time $\tau_\ast$ in Eq.(1) defines the ratio 
of the magnetic fields in two zones 
needed for an explanation of the flux ratio 
$J_1/J_2\simeq 1$. Indeed an effective modification of the injection spectrum
of radio electrons in the GeV region implies that
at these energies the overall escape time $\tau_{\rm esc}(E) \simeq 
\tau_{\ast}(E/E_\ast)^\delta$, and that it is significantly 
smaller than $t_0\simeq 300\,\rm yr$.  
The leakage of the electrons from $V_1$ will
result in $N_{2}(E) 
\sim N_1(E)\,t_0/\tau_{\rm esc}(E)$ electrons accumulated 
in the surrounding zone 2.
Using also Eq.(3), we then find   
\begin{equation}
\frac{B_1}{B_2} \simeq \left( 
\frac{t_0}{\tau_\ast}\right)^{\frac{2}{1+\beta_{2} }}
\left( \frac{2\,B_1}{1\,
{\rm mG}}\right)^{\frac{\beta_{2} -\beta_{1} }{1+\beta_{2} }}
\left( \frac{J_1}{J_2}\right)^{\frac{2}{1+\beta_{2} }} \; ,
\end{equation}
Thus, for $\tau_{\ast}\leq 0.1\,t_0$ the ratio 
$B_1/B_2\geq 3$.

In Fig.3  we show the energy distributions of electrons 
$n_{1,2}(E)= N_{1,2}(E,t_0)/V_{1,2}$ calculated the injection
spectrum with  
$\beta_{\rm acc} =2.24$ and $E_{\rm c}=35\,\rm TeV$,   
 and the escape time parameters  
$\tau_{\ast}=25\,\rm yr$ and $\delta =0.6$. 
At GeV energies the power law index of electrons in zone 
1 is indeed strongly modified, almost reaching the maximum theoretical
value $\beta_{1}  \approx \beta_{\rm acc}  +\delta$. 
However both at  lower and higher energies there is 
some flattening of $n_1(E) $.
At energies $E\leq 100\,\rm MeV$ the flattening is connected with the 
increase of the escape time to $\tau_{\rm dif} >100\,\rm yr$, which 
becomes comparable to the age of Cas~A, reducing thus the efficiency
of the ecape. At high energies   
$\tau_{\rm dif}(E)$ is limited by the energy-independent minimum escape time 
$\tau_{\rm min}$ in Eq.(16). This effect can be seen in Fig.3 for the 
dot-dashed curve at $E\geq 100\,\rm GeV$. Note, however, that  
for the  self-consistently calculated spectrum (solid curve)
of electrons in zone 1 the flattening is noticeable already at energies 
$E\geq 10\,\rm GeV$ where $n_1(E)$ becomes comparable to $n_2(E)$. At these
energies the gradients in the spatial distribution of   
relativistic electrons in the source become small, reducing     
the net exchange of particles between the zones, which again 
results in a less efficient modification of $Q(E)$. 

\begin{figure}[htbp]
\resizebox{8.8cm}{!}{\includegraphics{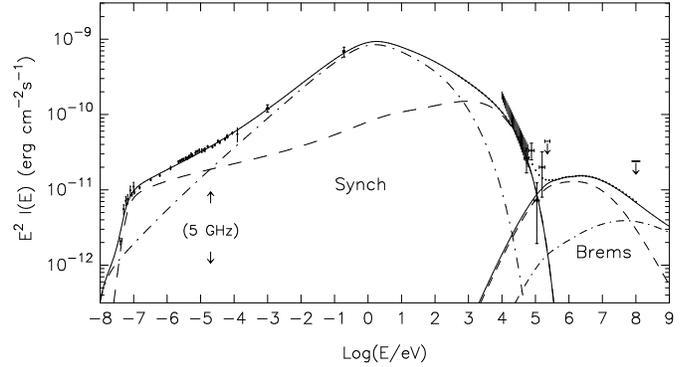}}
\caption{
The fluxes of synchrotron (heavy lines) and bremsstrahlung 
(thin lines) photons produced by electrons shown in Fig.3. 
The dashed and dot-dashed
curves correspond to the fluxes expected from the compact radio structures 
and the diffuse plateau, respectively, and the solid curves correspond to 
the synchrotron and bremsstrahlung fluxes of two zones. The full dots 
show the total nonthermal flux.}
\end{figure}

\begin{figure}[htbp]
\hspace{10mm}\resizebox{6.5cm}{!}{\includegraphics{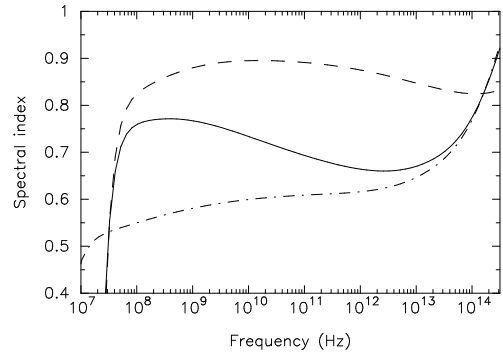}}
\caption{The spectral indices of the synchrotron fluxes corresponding
to the total (solid), zone 1 (dashed), and zone 2 (dot-dashed) radiations
shown in Fig.4} 
\end{figure}

 The synchrotron and bremsstrahlung fluxes are presented 
in Fig.\,4, and Fig. 5 shows the spectral indices 
at different frequencies.  
Because the  
synchrotron radiation in zone 1 is  
steep, it becomes dominant at low frequencies. The compactness of zone 1  
components 
leads to synchrotron self absorption at $\nu \leq 20\,\rm MHz$.
At frequencies above 5\,GHz the plateau emission dominates resulting in 
a noticeable flattening of the overall spectrum in the far infrared (FIR).
All these results are in agreement with expectations derived qualitatively
in Sect. 2.

For the model parameters used in Fig.4 the  energy content in relativistic
electrons in zones 1 and 2 is $W_{\rm e1}=1.3 \times 10^{48}\,\rm erg\,$ 
and $\,W_{\rm e2}=1.4 \times 10^{48}\,\rm erg$, respectively, and the energy 
in the magnetic fields  $W_{\rm B1}=2.6\times 10^{47}\,\rm erg$ and 
$W_{\rm B2}=6.8\times 10^{48}\,\rm erg$.  The total energy in relativistic 
electrons, $W_{\rm e}= W_{\rm e1}+W_{\rm e2}$, can be estimated analytically 
taking into account that: (a) 
the shape of the overall
energy distribution of electrons $N(E)= N_1(E)+N_2(E)$ repeats the power law 
injection $\propto E^{-\beta_{\rm acc} }$ until the break energy 
$\sim \! 300\,\rm GeV$ due to synchrotron losses (in zone 2), 
and (b) the total number of GeV electrons $N_\ast=E_\ast \,N(E_\ast)$ (with 
$E_\ast = 1\,{\rm GeV}\approx 2000\,m_{\rm e} c^2$) is 
defined mainly by electrons in zone 2, $N_\ast \approx N_{\ast 2}$,
which can be estimated using Eq.(3). For $\beta_{\rm acc} \geq 2.1$ 
we find 
\begin{equation}    
W_{\rm e} \approx \, 2.2\times 10^{46}
\;\frac{(2000)^{\beta_{\rm acc} -2}}{\beta_{\rm acc}  -2}
\left(\frac{B_2}{1\,\rm mG}\right)^{-\frac{\beta_{\rm acc} +1}{2}}\;\rm erg\; .
\end{equation}
For $\beta_{\rm acc} =2.24$ and $B_2=0.4\,\rm mG$ this equation results in 
$W_{\rm e}\approx 2.5\times 10^{48}\,\rm erg$, in agreement with the 
numerical results.

\begin{figure}[htbp]
\resizebox{8.7cm}{!}{\includegraphics{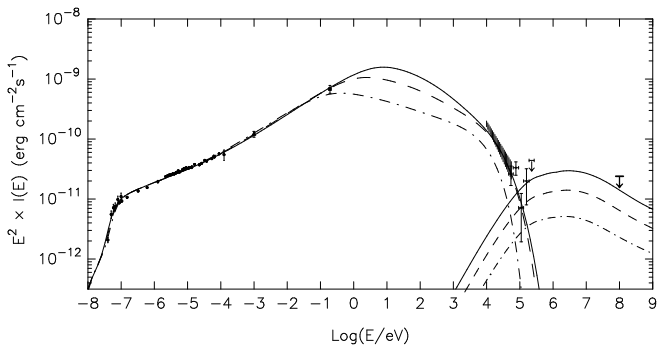}}
\caption{Radiation fluxes calculated for $\beta_{\rm acc}=2.2$
and 3 different magnetic fields 
$B_1$ in zone 1,  
$B_1=1\,\rm mG$ (solid), 1.6\,mG (dashed), and 3\,mG (dot-dashed),
assuming the same ratio $B_1/B_2=4$.
Exponential cutoff energies $E_{\rm c}$ and escape times $\tau_{\ast}$
in these 3 cases are:  $E_{\rm c}= 16\,\rm TeV$ and $\tau_{\ast}=26\,\rm yr$
(solid), $E_{\rm c}= 27\,\rm TeV$ and $\tau_{\ast}=20\,\rm yr$ (dashed), and 
$E_{\rm c}= 200\,\rm TeV$ and $\tau_{\ast}=14\,\rm yr$ 
(dot-dashed). For the assumed here number of zone 1 components  $K=150$
with radius $R_1=0.06\,\rm pc$ the sky covering fraction of the zone 1 
is 0.09\,.}
\end{figure}

In Fig.6 we show the fluxes calculated for $\beta_{\rm acc}=2.2$
and different
values of the magnetic fields in zones 1 and 2. For all 3 cases the spectra
of synchrotron radiation coincide up to the IR, but 
at higher frequencies they differ because of radiative break  
induced by synchrotron losses. Note that for 
interpretation of the X-ray fluxes above 10\,keV one needs  
higher values of the exponential cutoff
energy for higher magnetic field $B_1$: $E_{\rm c}=16\,\rm TeV$ for
 $B_1=1\,\rm mG$ and $E_{\rm c}=27\,\rm TeV$ for $B_1=1.6\,\rm mG$. 
For $B_1=3\,\rm mG$ the calculated synchrotron fluxes remain 
significantly below the observational data even assuming 
$E_{\rm c}=200\,\rm TeV$.

However, the synchrotron losses in high magnetic fields
effectively limit the maximum energy of 
electrons to values well below $E_{\rm c} \simeq 200\,\rm TeV$. 
For example, in the case of diffusive shock acceleration, 
the acceleration rate can be estimated as 
 ${\rm d}E/{\rm d} t \sim E u_{2}^2/D(E)$, where $D(E)$ is the diffusion 
coefficient and $u_2 $ is the speed of CBSR components with respect to 
the surrounding 
plasma. Equating this rate in the Bohm diffusion limit 
to the synchrotron loss rate, the characteristic maximum  energy of 
electrons can be estimated as:
\begin{equation}
 E_{\rm c}^{\rm (max)} \simeq 3\times 10^{13} 
\left(\frac{u_2}{3000\,\rm km/s}\right) \left(
\frac{B}{1\,\rm mG}\right)^{-1/2}\,\rm eV.
\end{equation}
Comparison with the values of $E_{\rm c}$ in Fig.~6 shows 
that diffusive shock acceleration could be consistent with synchrotron 
origin of the hard X-rays from Cas~A for  $B_1 < 2\,\rm mG$. 
Higher values of $B_1$ would require  a more efficient acceleration mechanism.
But anyway, the value of $B_1 = 3\,\rm mG$ is a very conservative 
upper limit of our model for the mean  
magnetic field in the bright radio structures if the measured hard 
X-ray flux has synchrotron origin. At the same time,
the lower limit for $B_1$ is about 1\,mG, which 
is needed for consistency of the bremsstrahlung  
with the flux upper limits of EGRET and OSSE detectors (see Fig.6). 
It is interesting that magnetic fields corresponding to the energy 
equipartition with relativistic electrons in the CBSR components 
are in the range $B_1\simeq (1.5-2)\,\rm mG$.

 Another feature to be addressed 
in this section is the wavelength-dependent rate 
$-{\rm d} \ln J(\nu)/{\rm d}t$ of the secular decline of  
radio fluxes. For calculations of this rate we need to specify the rates 
 of the magnetic field decline and the adiabatic losses of electrons.
For a spherical source (shell) homogeneously expanding with  speed $u_0$,
 the adiabatic losses  
$P_{\rm ad}=E/t_{\rm ad2}\,$,
where $t_{\rm ad2}\simeq t_{\rm exp} = R_0/u_0 \simeq 1000\,\rm yr$ 
corresponds to the current expansion time (`age') of the outer shell. 
In the `standard' power-law approximation for the magnetic field declining
with the shell radius as $B(R_0)\propto R_{0}^{-m}$, the characteristic
decline time 
$t_{\rm B}=B/(-{\rm d}B/{\rm d}t)=t_{\rm exp}/m$.  
For a magnetic field `frozen' into the spherically expanding source $m=2$, 
but $m<2$ if the  fields are effectively created in the shell. 
The assumption $m\simeq 1.5$ results in $t_{\rm B2} \simeq 700\,\rm yr$.
Concerning zone 1, we  note that 
along with the compact radio structures with declining brightness 
observations show also a large number of brightening components, and the 
timescales of the flux variations in the individual knots can be as small 
as few tens of years (Tuffs \cite{tuffs86}, AR96).
An estimate of the average time $t_{\rm B1}$ of the magnetic field decrease 
in CBSR components is thus problematic. Therefore we consider 
$t_{\rm B1}$ as a free model parameter, and neglect adiabatic losses
 of electrons there (if any -- especially as they are
supposed to be the sites of effective acceleration). 

\begin{figure}[htbp]
\hspace{7mm}\resizebox{7.cm}{!}{\includegraphics{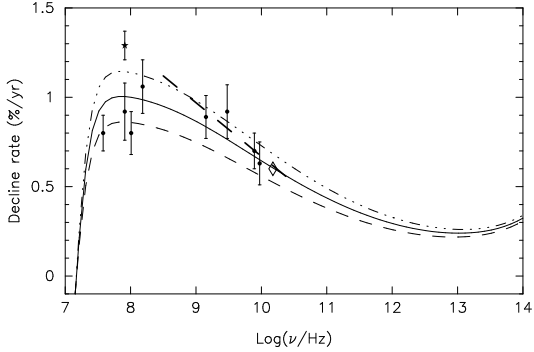}}
\caption{The rate of secular decline of the total flux of synchrotron
radiation shown in Fig.4, calculated for 
characteristic timescales (see text) $t_{\rm B2}=650\,\rm yr$ and  
$t_{\rm ad2}=1000\,\rm yr$ for zone 2, and different timescales for  
the magnetic field decline in zone 1:
  $t_{\rm B1}=115\,\rm yr$ (dot-dashed), 130\,yr (solid),
  150\,yr (dashed). The diamond shows the recent result of
O'Sullivan \& Green (\cite{osullivan}) at 15 GHz, all other data points are from
Rees (\cite{rees}). The heavy dashed straight line corresponds to
the earlier `best fit' estimate (see text).}
\end{figure}

Calculations of the secular decline rate of the fluxes 
at different frequencies are presented in Fig.~7. The straight
heavy dashed line shown corresponds to the  
decline rate deduced by Dent et al. (\cite{dent}) and Baars et al. (\cite{baars}) 
from comparison of the decline rate $\simeq 1.29\,\%/ \rm yr$ 
measured then at 81\,MHz by Scott et al. (\cite{scott69}) 
with the data in the 1-10\,GHz region. Later measurements 
have shown that the decline rate around $\nu \sim 100 \,\rm MHz$
is at the same level $\sim (0.8-1)\,\% /\rm yr$ as at 
$\nu \sim 1\,\rm GHz$, which has placed doubt on the frequency dependence 
of the decline rate in general
(Rees \cite{rees}; Hook et al. \cite{hook}). 
However, the recent result of O'Sullivan \&
Green (\cite{osullivan}) that the flux decline rate at 15\,GHz is about 
$0.6\%/\,\rm yr$,
seems to confirm the effect of slower decline of radio fluxes at high 
frequencies.

Figure 7 shows that practically all the observational data can be well 
explained assuming the average magnetic field decline time in zone 1 
of order $t_{\rm B\,1} \sim 120-150\,\rm yr$.
 Note that a significant drop 
of the  secular decline  rate predicted at $\nu < 30\,\rm MHz$ is connected
with the gradual decrease of synchrotron self-absorption in zone 1.

\section{CBSR components as acceleration-passive sites}

Consider now the hypothesis that acceleration of electrons takes place
only in zone 2 and is negligible in zone 1. 
In this case the source function $Q_2(E) $ for 
zone 2 is in the form of Eq.(18), while  
$Q_1(E)$ for zone 1 is defined only by the rate 
$Q_{2\,1}(E)$ of advection of relativistic electrons from the  
plateau region.  The required for zone 1  
relativistic electron density enhancement could be then 
supposed as being due to adiabatic compression
of the zone 2 electrons on the front surface $S_{-}$ of fastly moving
CBSR components, and their subsequent accumulation in zone 1 over timescales 
$\Delta t\sim \tau_{\rm esc}$. Neglecting in Eq.(7) the diffusion term 
on the surface $S_{-}$, in compliance with the assumption of
negligible diffusive acceleration there, and retaining only the convection 
and the adiabatic compression terms, the equation describing the transport
of the electrons through $S_{-}$ reads: 
\begin{equation}
u_x\frac{\partial f_{p}}{\partial x} - 
\frac{\partial u_x}{\partial x}\, 
\frac{p}{3}\frac{ \partial f_{p}}{\partial p} =0\; ,
\end{equation}
where $f_{p}\equiv f_{p}(x,p,t)$ is the momentum distribution function,
$x$ is the spatial coordinate and $u_x$ is the plasma speed 
along the $x$-axis perpendicular to the 
compression layer. General solution to this equation reads: 
$f_p(p,x)=F_0[p (u_x/u_0)^{1/3}]$, where  
$F_0(p) = f_{p}(p,x_0)$ is an arbitrary initial function. 
Connecting the electron distribution functions 
across the layer, the  
injection rate 
 of relativistic electrons from zone 2 into zone 1 is found:
\begin{equation}
 Q_{2\,1}(E)\simeq N_2(E \,\rho_{\rm c}^{-1/3}) \, \rho_{\rm c}^{-1/3}
\, S_{-} \, u_2 \,V_{2}^{-1} \; ,
\end{equation}
where $\rho_{\rm c}=u_2/u_1$ is the mean compression ratio,
and $u_2$ and $u_1$ correspond to the plasma speeds of the plasma 
in the upstream and downstream regions in the rest frame of the 
CBSR components.

Considering possible values of $u_2$, note that for the 
mean radial expansion age of radio knots $t_{\rm exp}=r/v_{\rm r}$ within
$(550-1000)\,\rm yr$ (Tuffs \cite{tuffs86}; Anderson \& Rudnick \cite{ar95}) the   
characteristic speed of the knots may be estimated as
$v_{\rm kn}\simeq (1700 -3200)\,\rm km/s$. Taking into account that 
surrounding plasma in the 
shell is moving in the same direction, and as evidenced by a shorter 
X-ray expansion time scales $\sim 600\,\rm yr$ 
(Koralesky et al. \cite{koralesky}; Vink et al. \cite{vink}), 
appears to be overtaking most of the compact
radio knots, the value $u_2\sim 2000 \,\rm km/s\,$ seems  
a reasonable estimate for the relative speed of radio knots with respect
to the surrounding medium. 
For the bright radio ring $t_{\rm exp}\simeq 950\,\rm yr$ 
(Tuffs \cite{tuffs86}), corresponding to $v_{\rm ring}\simeq 1700\,\rm km/s$,
while the speed of freely expanding ejecta upstream 
of the reverse shock, i.e. at $r < R_{\rm ring}$, is about $(5000-5300) 
\,\rm km/s$. Thus, for the radio ring   
$u_2\sim 3500\,\rm km/s\,$, and we can take $3000 \,\rm km/s\,$
as a reasonable maximum value of $u_2$ for the zone 1. 

The compression ratio $\rho_{\rm c}$ can be approximated as the ratio
of magnetic fields  $\rho_{\rm c}\simeq B_1/B_2$.     
The energy distributions of the radio electrons  
can be connected as
\begin{equation}
\frac{N_1(E_\ast)}{N_2(E_\ast)} \simeq \frac{V_1}{V_2}\,
\left( \frac{B_1}{B_2} \right)^{\frac{\beta_{2}  -1}{3}}
\frac{u_2\,\tau_\ast}{2\,R_1}\; ,    
\end{equation}
 using the relation 
$N_1(E_\ast)\simeq Q_{2\,1}\times \tau_\ast$. Eqs.~(3) and (24)
result in 
\begin{equation}
\left( \frac{B_1}{B_2}\right)^{\frac{5\,\beta_{2}  +1}{6}} \simeq \,
\frac{J_1}{J_2}\,\frac{2\,R_1\,V_2}{u_2\,\tau_\ast\,V_1}\,
\left( \frac{2\,B_{1}}{1 \rm mG}\right)^{\frac{-\Delta \beta}{2}}\; ,
\end{equation}
where $\Delta \beta = \beta_{1} -\beta_{2} $.
Since  the volume filling factor of the compact zone 1 structures is   
$V_1/V_2\sim 0.5\,\%$ or less, and since  for effective modification of the
electron spectra by escape one needs
$\tau_{\ast} \leq 0.1 \,t_0=30\,\rm yr$, the ratio of the magnetic
fields in the two zones should be rather high,  $B_1/B_2 > 10$.  

The results of numerical calculations are shown in Fig.8.
The total flux seems to fit the data practically equally well as in Fig.\,4
earlier. However, a closer look to these figures reveals  significant
differences. First of all,  in Fig.\,8 the agreement with the 
radio data is reached 
due to a much higher contribution of zone 1 to the overall flux. 
In particular, at 5\,GHz the partition of the fluxes corresponds to $J_1/J_2
\simeq 3$, instead of $J_1/J_2\simeq 1 $. 
Secondly, although here we have assumed high $\delta=0.75$, the 
calculations show that the spectral index
of the electrons swept up from zone 2,
$\beta_2=2.2$, was steepened in zone 1 only to values 
$\beta_1\leq 2.6$, well below the `expected' value 2.95. Obviously, this is
far insufficient for explanation of the observed steep spectral indices of the 
radio knots.

\begin{figure}[htbp]
\resizebox{8.8cm}{!}{\includegraphics{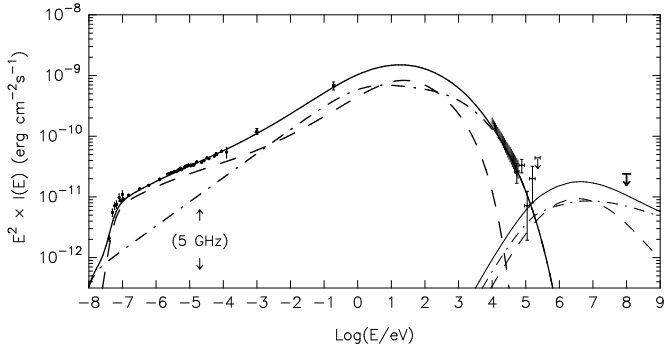}}
\caption{Radiation fluxes expected in the scenario when acceleration occurs 
only in the zone 2 (diffuse plateau), but not in the compact  
zone 1. The solid lines show the total fluxes, while the dashed and dot-dashed
lines correspond to the contributions from the zone 1 and zone 2, respectively.
Model parameters are: $B_1= 3.8\,\rm mG$, $B_2= 0.2\,\rm mG$
$\beta_{\rm acc}  = 2.2$, $E_{\rm c} =30,\rm TeV$, 
$\tau_{\rm ast}= 20\,\rm yr$, 
$\delta =0.7$, $K=200$, $R_1=0.06\,\rm pc$, and $u_2=3000\,\rm km/s$.}
\end{figure}

\begin{figure}[htbp]
\hspace{10mm}\resizebox{6.5cm}{!}{\includegraphics{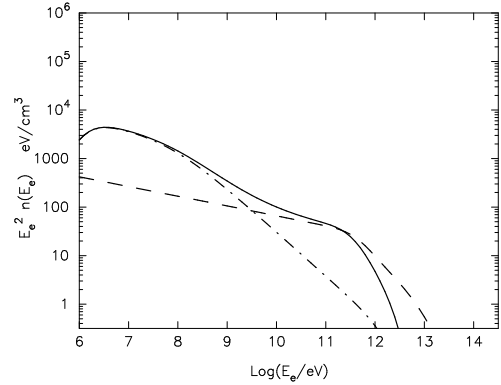}}
\caption{The energy distributions of the electrons formed in zone 1 
(solid line) and
zone 2 (dashed line) in the framework of `adiabatic compression' scenario for
zone 1 electrons, calculated for parameters in Fig.8.
The dot-dashed curve corresponds to the 
the energy distribution of electrons in zone 1
when the return diffusive flux of electrons from zone 2 into zone 1 is
neglected.}
\end{figure}

The reason for  difficulties arising in the scenario that assumes only 
passive compression, but not active acceleration of electrons in zone 1, 
can be understood from Fig.\,9 where we show the spatial densities
 $n_{1,2}(E)$ of 
the electrons in zones 1 and 2. 
The dot-dashed line shows the spectrum of the 
electrons which would be formed in the zone 1 if one neglects the 
term $\propto n_2/\tau_{\rm dif}$ in Eq.(14). This spectrum is steep.
But when the
effect of spatial density gradients  between the two zones is 
taken into account correctly, the situation dramatically changes. The
energy dependent propagation can result in a significant steepening 
of the source spectra  only in the regions of higher spatial densities
 of relativistic particles. 
Otherwise diffusive exchange of particles
tends to equalize $n_1(E)$ with $n_2(E)$ resulting in
$\beta_{1} \rightarrow \beta_{2} $ independently of $\delta$, until
the radiative losses become important.

Thus, for an effective modifications of the initial source spectrum
of the radio electrons the condition $n_1(E)\gg n_2(E)$ should be satisfied. 
The ratio of electron densities expected at 1\,GeV can 
be estimated from Eqs.~(25) and (26) (assuming $\beta_{2} \simeq 2.2$) as:
\begin{equation}
\frac{n_{\ast 1}}{n_{\ast 2}} \simeq 
\left( \frac{J_1}{J_2}\right)^{0.2}
\left( \frac{V_2}{V_1}\right)^{0.2}   
\left(\frac{u_2\,\tau_\ast}{2\,R_1}\right)^{0.8}
\left( \frac{2\,B_{1}}{1 \rm mG}\right)^{-0.1\,\Delta \beta} .
\end{equation}
For the parameters used in Fig.8 this equation predicts  
$n_{\ast 1}/n_{\ast 2}\approx 2$,  in agreement with numerical calculations, 
which is smaller by a factor of $15$ than similar ratio in Fig.3. 
Calculations show that it is not easy to increase this ratio significantly, 
remaining within the `adiabatic compression' scenario.

Thus, a self-consistent explanation of 
the observed radio fluxes of Cas~A is very problematic, unless 
we assume an 
effective acceleration of electrons in the CBSR  
components  in order to build up sizeable gradients in the spatial 
distribution of the radio electrons in the source.

\section{Results}

The spatially non-uniform source model, even in its 
simplest two-zone form, allows us to unify into a single 
picture many observational data on the broad-band nonthermal radiation of Cas~A.
The following is the summary of observational features of Cas~A which could be 
explained in the framework of this model. 
\vspace{2mm}

\noindent
The brighter radio structures tend to be steeper (AR96) to the extent that the 
energy density of relativistic electrons there would be higher. 
In principle, even assuming the same hard power law index  for the acceleration, 
e.g $\beta_{\rm acc}\simeq 2.2$, and $\delta\simeq 0.6$ for the escape of
the electrons, it is possible to explain the observed variations of the 
spectral indices of individual compact radio structures 
from $\alpha_{2} \sim 0.6$ to $\alpha_{1} \sim 0.9$.
Such variations can be connected with two effects. 
\vspace{1mm}

\noindent
({\sc a})~~ The intrinsic index of some knots will be flatter 
than the maximum possible $\alpha_{1} = \alpha_{2} + \delta/2$ if the local
gradient $n_1/n_2$ is not sufficiently high, and if the characteristic
escape time $\tau_{\ast}$ of GeV electrons from these knots is larger 
than $\sim 30\,\rm yr$. The impact of the latter effect can be seen in 
Figs. 10 and 11, where we show the fluxes and spectral indices calculated
in the modified two-zone model approach, subdividing the CBSR structures
in two groups and assuming 2 different spatial/temporal scales for larger 
(`ring') and smaller (`knot') components in zone 1.

\begin{figure*}[htbp]
\hspace{1.6cm}\resizebox{12.6cm}{!}{\includegraphics{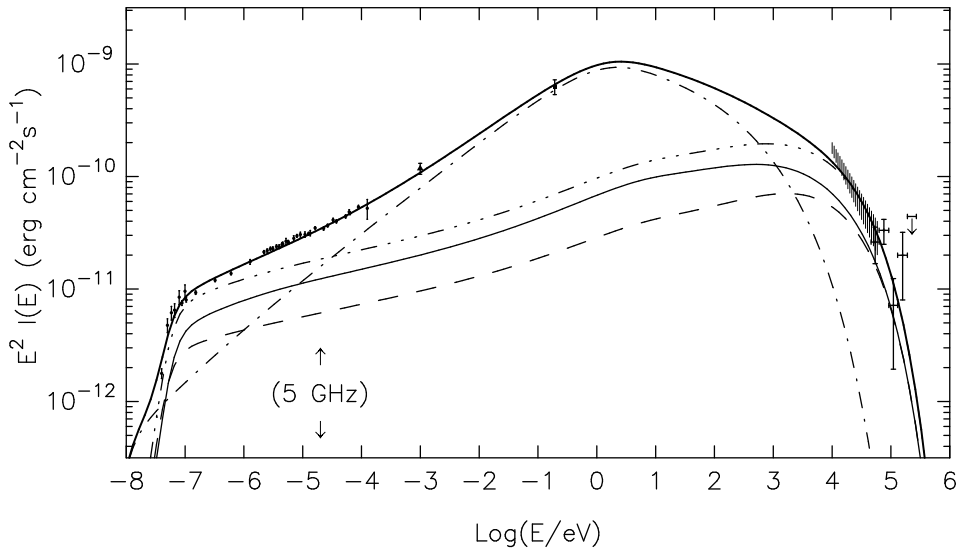}}
\caption{Fluxes expected in the framework of the modified  
two zone model, when in zone 1 more compact radio knots (dashed line)
and larger fragments of the radio ring 
(thin solid line) are separated. The 3-dot--dashed line 
shows the total flux from zone 1, and the dot-dashed line shows 
the flux from zone 2. The total synchrotron flux   
is shown by the heavy solid line. 
The model parameters are the following. For the knots: 
 $R_{\rm 1. kn}=0.033\,\rm pc$, $\tau_{\ast }= 15\,\rm yr$,
$\delta=0.6$,  $K_{\rm kn}=180$;
for the ring fragments $R_{\rm 1. ring}=0.1\,\rm pc$, 
$\tau_{\ast }= 35\,\rm yr$, $\delta=0.6$,  $K_{\rm ring}=20$. 
The same $\beta_{\rm acc}  = 2.2$ and 
cutoff energy $E_{\rm c} = 35\,\rm TeV$ for the accelerated
electrons,  and magnetic fields $B_{1.\rm kn}= B_{\rm 1. ring} = 
1.5\,\rm mG$ in the knots and the ring is assumed.
The magnetic field $B_2= 0.36\,\rm mG$.} 
\end{figure*}

\begin{figure}[htbp]
\hspace{5mm}\resizebox{7.5cm}{!}{\includegraphics{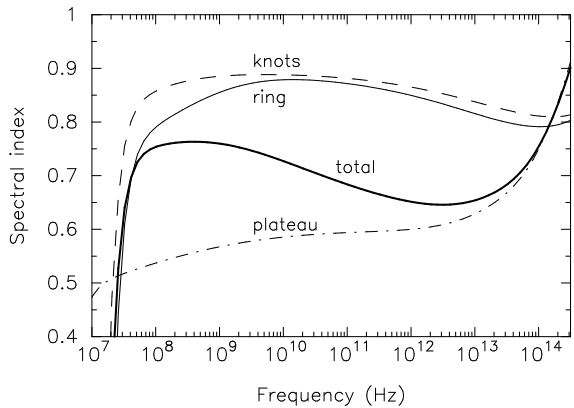}}
\caption{The spectral indices of the radiation components shown in Fig.10}
\end{figure}

\vspace{1mm}

\noindent
({\sc b})~~The intrinsic spectrum of individual knots could be steep, up to 
$\alpha_{1}\simeq 0.9$, but the observed spectrum may be flatter, 
depending on the degree of flux contamination 
by the flat-spectrum radiation from the diffuse plateau along the line of sight 
to the knot. This effect may be relevant mainly to those cases when 
the background subtraction is problematic (e.g., for the structures with
insufficiently high contrast to the plateau).

\vspace{1mm}

\noindent
The strong correlation between the spectral index  
and the projected position of radio knots observed by AR96 could be connected 
with both effects ({\sc a}) and ({\sc b}). Indeed, the 
brightness of the diffuse plateau seems to decrease from the radio
ring  towards the outer edge of the shell more rapidly than one would
expect merely on the base of the projection effect in a spherical geometry
(see Fig.7 from ARLPB). 
This implies that the 
density $n_2$ of radio electrons could be  
higher at distances closer to the ring. Then 
the steepest knots would be found predominantly  
closer to the edge of the plateau, because (i) the local  ratio 
$n_1/n_2$ might be higher far from the reverse shock, and (ii) contribution
of the flat spectrum plateau emission to the steep spectrum flux of the knots
is smaller in directions to the projected periphery of the shell. 
     
\vspace{1mm}

\noindent
The cutoff of the radio spectrum below 20\,MHz is 
explained by the synchrotron self-absorption of the flux 
of zone 1.  This interpretation is possible basically
because the model predicts that the intrinsic fluxes of the 
compact bright structures are much steeper than the spectra in the diffuse
plateau. Then  at low frequencies the flux of CBSR components should dominate 
the total radio emission. Therefore synchrotron self-absorption of 
only this flux, which becomes possible due to high density of particles
(and fields) in those compact structures, is sufficient to comply with the data.

 It is worth noticing that due to variations of the physical parameters in the
individual CBSR components contributing to zone 1 flux, one 
could generally expect that the synchrotron 
absorption of that flux would be in reality  smoother than 
it is shown in Fig.~4 where we have assumed the same parameters for 
all counterparts of zone 1. The position of the 
characteristic cutoff frequency for an individual component $j$ is found from 
the condition $\tau_j(\nu)=1$ where $\tau_j=R_j \kappa(\nu)$ is its opacity. 
From Eqs. (3) and (4) it follows that $\tau_{j}(\nu)\propto A_j 
\nu^{-(2+\beta_j/2)}$, with $A_j=R_{j}^{-2} f_j B_{j}^{0.5}$ where
$f_j$ is the knot flux (luminosity) at some fixed frequency (say 1\,GHz). 
Because of the very strong dependence of $\tau_j$ on $\nu$, a significant
broadening of the synchrotron turnover frequency for the ensemble of CBSR
components (say by a factor 2) would be expected only if  the 
 {\it luminosity-weighted} dispersion $\sigma$ 
of the parameter $A_j$ is very large ($\sigma\simeq 10$). 
The fact that the observed turnover is sharp may
then imply a quite reasonable possibility -- that the  
distribution of the parameters $\{A_j\}$ is not far from  
 `Gaussian noise'. In that case one would normally
expect $\sigma\sim 1$ (or perhaps even less, given the expected correlation
of $f_j$ and $R_j$), so the synchrotron
turnover position would be effectively dispersed, or broadened, only
by a factor roughly $(1+\sigma)^{0.3}\simeq 1.2$, i.e. about $\pm 20\%$
around the mean position. Note that the overall flux in Fig.10, where 2 
significantly different sizes for the large and small components are assumed,
show practically the same sharp cutoff below 20\,MHz as in Fig.4.       

\vspace{1mm}

\noindent
Observations at low frequencies  by KPDE, at a resolution of about 
$20^{\prime \prime}$, show that the spectral index of Cas~A between
333\,MHz and 1.38\,GHz is approximately constant, $\alpha \simeq 0.75$,  
at angular radii $\theta \leq 100^{\prime\prime}$, but that 
it is quickly increasing
to $\alpha \leq 0.95$ as $\theta \rightarrow 150^{\prime\prime}$. 
In principle, this effect might be connected with the steep spectrum radio 
knots which have been detected by AR96 but cannot be resolved in the low 
frequency maps of KPDE. Because of the rapid decline of the plateau brightness
 at large angular distances from the ring, the CBSR components (the 
radio knots and bow shocks)   
increasingly contribute to the `diffuse' overal flux closer to the periphery. 
We do not exclude that perhaps in the framework 
of a more spatially structured model (that would allow large scale inhomogeneities
in the plateau region itself) one could expect also 
some steepening of the intrinsic diffuse flux. However, given much slower and 
less efficient spectral modifications for the large scale inhomogeneities, 
the propagation effects alone would be able to explain only rather 
moderate steepening of the intrinsic plateau emission. If the acceleration
is efficient so that the primary spectrum is really hard,  the steepening to 
values $\alpha\sim 0.9$ seems to require a dominating contribution of the
 compact structures in the overall flux detected by KPDE.

\vspace{1mm}

The two-zone model suggests 
a possible interpretation for the `discrepancy' 
between the spectral index $\alpha = (0.7-0.75)$ found by KPDE
at the radius of the bright radio ring, 
and observations of WDS90 who 
found $\alpha \simeq 0.83$ if  {\it only} the bright
 components at approximately the same radial distances are considered. 
Both results might be understood if we take into account that the 
spectral index found by KPDE 
corresponds to the flux of  two zones integrated along the 
geometrical circles of different fixed radii, and hence would be more
affected by the flat plateau flux than the result 
found in  WDS90 (compare the heavy and thin solid lines
in Fig.11).

\vspace{1mm}

\noindent
The flattening of the radiation spectrum observed at millimeter 
 wavelengths (Mezger et al. \cite{mezger}) 
is a natural consequence of the `flat + steep' representation of the
total flux in the two-component approach, with approximately equal 
contributions from these components around 5\,GHz. The two-zone model
also predicts that the flux measured by Tuffs et al. (\cite{tuffs97}) 
around $6\,\mu m$
should be predominantly synchrotron in origin.   

\vspace{1mm}

\noindent
The unusual dependence of the secular decline of radio fluxes 
on frequency, which is at a constant level 
at $\nu\sim 40\,{\rm MHz}- 1\,\rm GHz$ 
but seems dropping at higher frequencies, 
is explained by varying contributions of the flat 
and steep spectrum components at different frequencies. 
This interpretation     
suggests that 
on average the net flux of the bright compact components in Cas~A 
drops faster than the diffuse plateau emission, in agreement with
observations of Tuffs (\cite{tuffs83}).

\vspace{1mm}

\noindent
The synchrotron origin of the X-rays above 10\,keV can be
explained if electrons in the compact CBSR components 
are accelerated to energies of few tens of TeV. These energies can be 
reached, e.g., by diffusive shock acceleration in the Bohm limit, 
which seems  seems a plausible mechanism for production of relativistic 
electrons at the reverse shock presumably connected with the bright
radio `ring'. 
In principle,  for the radio knots as well the diffusive 
acceleration at the bow shocks could result in the high electron energies 
needed for X-ray emission. 
Perhaps, another possibility for acceleration of electrons in 
the radio knots could be connected with reconnection of the magnetic
field lines strongly amplified, as shown by Jones et al. (\cite{jones94}),
in thin turbulent layers behind the bow shocks of dense gas `bullets'
at the stage of their fast disruption 
by Rayleigh-Taylor and  Kelvin-Helmholtz instabilities. 

\vspace{1mm}

\noindent
Hard spectral indices for accelerated particles weaken 
the constraints on the energetics of relativistic electrons
in Cas~A. For example, in the case of $\beta_{\rm acc}=2.24$, and 
$B_1=1.5\times 10^{-3}\,\rm G$ and $B_2 = 4\times 10^{-4}\,\rm G$ used
in Fig.4, the total energy $W_{\rm e} = 2.7\times 10^{48}\,\rm erg$.
The energy $W_{\rm e}$ needed in the case of $\beta_0=2.54$ for  
a  uniform shell with the same magnetic field
$B_0=B_2$ is by a factor of 10 larger.

\section{Conclusions}

The radio fluxes of the prototype young SNR Cas~A show a large variety of 
spectral and temporal features which are very difficult 
to incorporate into one self-consistent 
picture, if one remains in the framework of a single-zone 
(i.e. uniform) model approach and thus has to attribute the spectral indices 
deduced from 
radio observations to the source spectra of accelerated electrons,   
since radiation losses cannot modify the energy distribution of 
radio electrons during the lifetime of Cas~A.  
The interpretation of the radio data essentially changes if  
we take into account that energy dependent propagation of relativistic 
particles is able to modify their energy distribution in a 
spatially inhomogeneous source on timescales much 
shorter than the radiative loss time. The efficiency and extent 
of these modifications depend on the strength of the   
gradients in the spatial distribution of the particles in the source.

The simplest spatially inhomogeneous model is the one where the radio 
source is subdivided into two zones 
with different energy distributions of relativistic
electrons and different magnetic fields.
The basic assumption of the model is that the number density of 
radio electrons in the compact bright radio components 
(predominantly included in zone 1)  is much higher than
in the surrounding diffuse shell between the reverse shock and the blast wave
(zone 2).
The fulfillment of this assumption is contingent on an efficient acceleration 
of electrons in these radio bright components.
 In principle, acceleration of 
electrons in the diffuse shell is not needed at all. Note, however,
that the model does not exclude a contribution from such acceleration 
provided that it is not so powerful as to  
noticeably reduce the gradients in the spatial density of electrons 
established by acceleration in the CBSR components, which would 
otherwise diminish 
the efficiency of spectral modifications in zone 1.

 The possibility for exchange of particles between these zones on an energy 
dependent timescale $\tau_{\rm esc} \propto E^{-\delta}$, with 
$\delta\sim 0.5-0.6$, allows us to suggest a single hard power law injection  
spectrum of electrons with an index $\beta_{\rm acc}  \simeq 2.2-2.3$,  
implying an efficient acceleration process.
Then, depending on the relativistic
electron density contrast and physical parameters of individual
component $j$, the leakage of electrons from those CBSR components can result
in the steepening of the injection spectrum up to $\beta_{1,j}\leq 2.9$,
in agreement with the maximal spectral indices of the knots 
$\alpha_{1,j}\leq 0.95$.  
The energy distribution of radio electrons $N_2(E)$
in the extended plateau will show the hard power-law index of injection, 
$\beta_{2} \approx\beta_{\rm acc} $. As a general remark we note that
energy-dependent propagation in any non-uniform medium would always tend
to flatten the energy distribution of particles in the regions with 
{\it lower} density, i.e. which are typically more extended.

\vspace{3mm}

\noindent
The model leads to a number of predictions.

\vspace{1mm}

The two-component decomposition of the overall flux, with
$J_1\sim J_2$ at 5\,GHz, predicts that the sites of current or very recent
acceleration will become more pronounced in the high resolution (few arcsec)
maps at lower frequencies, where  
 the contribution of the diffuse plateau to the overall flux will decrease.
This should result in a significant increase of the brightness contrast
for the remaining compact structures (which belong to zone 1), reaching its 
maximum around 40\,MHz (i.e. at the minimum frequencies not affected by
synchrotron self-absorption).
\vspace{1mm}

At frequencies $\leq 30 \,\rm MHz$ 
the bright compact structures will quickly 
disappear, but the plateau emission, in the form of a
weak diffuse shell, may become dominant again. 
This shell would probably have an  apparent radius less than 
$150^{\prime\prime}$ if the
emissivity of the shell is indeed strongly decreasing towards the 
blast wave, implying very low brightness at large radii.

\vspace{1mm}

Correlations between the spectral index and the geometrical 
thickness of the radio knots of similar brightness 
could be expected on the high resolution maps of Cas~A taken at 1.4-5\,GHz.
The radio knots with a smaller thickness, but the same brightness, 
would tend to be steeper since the escape times 
in those knots could be smaller. Note however that the escape time is not the 
only parameter affecting the efficiency of spectral modifications. 
In particular, the `steepness-compactness' correlation could be different  
for knots in the fading stage (which are `older', with declining 
injection of new particles) and in the brightening stage. Besides, such a 
correlation could be significantly affected by 
different magnetic fields in the knots.  Nevertheless, the search for 
correlations between spectral index and compactness of the knots seems 
worthwhile.

\vspace{1mm}

At frequencies  above 10\,GHz the brightness contrast will 
decrease, and the CBSR structures may appear less pronounced. 
The total emission will be dominated by the diffuse flat-spectrum plateau, 
and the spectral index may decrease to $\alpha\approx 0.65$ (see Fig.11).
The structure of zone 2, in particular the outer edge of the
shell, will be better discernible.

\vspace{1mm}

An important prediction concerns  the 
character of the long term evolution of the radio pattern of Cas~A. 
One can expect that during the next $\Delta t \sim t_{\rm B1}\sim 
(150 - 200)\,\rm yr$, when the total emission will become dominated 
by the diffuse plateau and the compact radio structures may become less 
common, the remnant will have a much more 
uniform radio shell with a spectral index $\sim 0.6$, and  
Cas~A may become similar to Tycho's SNR\footnote{An alternative 
explanation for the contrasting brightness, spectral 
and morphological characteristics of Tycho's SNR could be that, 
as the remnant of a type Ia event, this source never contained compact
efficient accelerators as appear to be now present in the ejecta - 
circumstellar interaction of Cas A, and that
all particle acceleration in Tycho's SNR has occurred at the blast wave.}
(e.g. see Klein et al. \cite{klein}).
Interestingly, the overall efficiency of electron acceleration
at that time might be even lower than at present, while the
radio spectrum will be significantly harder reflecting the spectrum of 
the acceleration which is taking place presently. 
More generally, the energy dependent propagation of radio electrons in 
a spatially inhomogeneous  medium  can explain the trend 
(see e.g. Green \cite{green}; Jones et al. \cite{jones98}) that young clumpy 
shell-type SNRs often 
exhibit radio spectra that are significantly steeper than those of older  
ones showing a typical spectral index $\alpha\sim 0.5$.  

\vspace{1mm}

 At X-ray frequencies the appearance of Cas~A may be very
different in the soft X-ray and hard X-ray domains. At photon energies 
below $10\,\rm keV$ the radiation is dominated by thermal emission
of the gas in the extended region from 
the reverse shock up to the blast wave. Apart from the thermal 
diffuse emission we could
expect a noticeable contribution of the nonthermal X-rays from compact 
radio structures. This radiation could be more easily distinguished 
from the thermal emission in the case of compact bright radio knots at the 
periphery of Cas~A, if the X-ray spectra would appear relatively flat in the 
1-5\,keV region (see Figs.~4 and 11) and the line emission would be deficient. 
These observations will be possible with the high spectral and 
angular resolutions of XMM and ASTRO-E. 

\vspace{1mm}

In hard X-rays above 10\,keV the appearance of Cas~A may
significantly change. In the case of a nonthermal origin of this radiation,
all acceleration sites of the highest energy electrons 
may become clearly visible. Note, however, 
that the angular resolution of the detectors needed to reveal significant
spatial changes at these energies must be better than 
$10^{\prime\prime}$.  
The observation of such hard nonthermal X-ray fluxes from the compact 
radio knots and radio ring would impose an upper limit on the magnetic field
in these structures $B_1< 3\,\rm mG$, with the probable value expected in the
range $B_1 \simeq (1-2)\,\rm mG$. The model predictions for 
magnetic fields in the diffuse shell are $B_2\simeq (0.3 - 0.5)\,\rm mG$.  

\vspace{1mm}

At energies above 100\,keV we predict very flat spectra
of the soft $\gamma$-ray  fluxes due to bremsstrahlung which perhaps
would be observable for  
the forthcoming ASTRO-E and INTEGRAL telescopes.

\vspace{1mm}
   
 For the high energy $\gamma$-rays, $E\geq 100\,\rm MeV$,  
fluxes at a level of 0.1-0.5 of the EGRET flux upper limits are expected. 
These fluxes should be easily detected by the GLAST instrument.

\vspace{3mm}
  
\noindent
In summary, the steepness of radio spectra of bright and compact structures 
in clumpy radio sources is not an indication of inefficient acceleration, 
but rather a natural consequence
of very efficient acceleration which  builds up high spatial gradients of 
relativistic electrons in those sources increasing the efficiency of 
spectral modifications due to their energy dependent escape.

\begin{acknowledgements}
The authors thank Larry Rudnick for very useful 
discussions, and the anonymous referee for very helpful comments. 
RJT thanks NRAO for hospitality during his visits to the VLA 
in 1984 and 1985 and to Rick Perley, Steve Gull and Martin Brown for 
support in the observations and data reduction which led to Fig.1.
The work of AMA was supported through the Verbundforschung
Astronomie/Astrophysik of the German BMBF under the grant No. 05-2HD66A(7).
\end{acknowledgements}


\begin{thebibliography}{}

\bibitem[1997]{allen}
Allen G.E., Keohane J.W., Gotthelf E.V., et al., 1997, ApJ 487,
L97


\bibitem[1995]{ar95}
Anderson M.C.,  Rudnick L., 1995, ApJ 441, 307
                                                           

\bibitem[1996]{ar96}
Anderson M.C.,  Rudnick L., 1996, ApJ 456, 234 (AR96)

\bibitem[1991]{arlpb}
Anderson M., Rudnick L., Leppik P., Perley R., Braun R.,
1991, ApJ 373, 146 (ARLPB)

\bibitem[1980]{ashworth}
Ashworth W.B., 1980, J. Hist. Astr. 11, 1

\bibitem[1999]{aa99}
Atoyan A.M., Aharonian F.A., 1999, MNRAS 302, 253

\bibitem[1977]{baars}
Baars J.W.M., Genzel R., Paulini-Toth I.I.K.,  Witzel A., 1977,
A\&A 61, 99

\bibitem[1979]{becker}
Becker R.H., Holt S.S., Smith B.W., et al., 1979, ApJ 234, L73


\bibitem[1977]{bell77}
Bell A.R., 1977, MNRAS 179, 573

\bibitem[1975]{bell75}
Bell A.R., Gull S.F.,  Kenderdine S., 1975, Nat 257, 463


\bibitem[1997]{berezhko}
Berezhko E.G.,  V\"olk H.J., 1997, Astroparticle Phys. 7, 183

\bibitem[1987]{braun}
Braun R., Gull S.F., Perley R.A., 1987, Nat 327, 395

\bibitem[1978]{chevkirsh}
Chevalier R.A.,  Kirshner R.P., 1978, ApJ 219, 931


\bibitem[1978]{chevetal}
Chevalier R.A., Oegerle W.R., Scott J.S., 1978, ApJ 222, 527

\bibitem[1984]{chini}
Chini R., Kreysa E., Mezger P.G., Gem\"{u}nd H.-P., 1984, A\&A 137, 117

\bibitem[1980]{cowsar80}
Cowsik R., Sarkar S., 1980, MNRAS 191, 855

\bibitem[1984]{cowsar84}
Cowsik R.,  Sarkar S., 1984, MNRAS 207, 745

\bibitem[1974]{dent}
Dent W.A., Aller H.D., Olsen E.T., 1974, ApJ 188, L11

\bibitem[1979]{dickel}
Dickel J.R.,  Greisen E.W., 1979, A\&A 75, 44

\bibitem[1991]{ellison}
Ellison D.C.,  Reynolds S.P., 1991, ApJ, 382, 242

\bibitem[1996]{esposito}
Esposito J.A., Hunter S.D., Kanbach G., Sreekumar P., 1996, ApJ 461,
820 

\bibitem[1980]{fabian}
Fabian A.C., Willingale R., Pye J.P., Murray S.S.,  Fabbiano G.,
1980, MNRAS 193, 175

\bibitem[1997]{favata}
Favata F., Vink J., Dal Fiume D., et al., 1997, A\&A 324, L49

\bibitem[1988]{fesen}
Fesen R.A., Becker R.H., Goodrich R.W., 1988, ApJ 329, L89 

\bibitem[1975]{fichtel}
Fichtel C.E., Hartman R.C., Kniffen D.A., et al., 1975, ApJ 189, 163

\bibitem[1988]{green}
Green D.A., 1988, Ap\&SS 148, 3 

\bibitem[1992]{hook}
Hook I.M., Duffet-Smith P.J., Shakeshaft J.R., 1992, A\&A 255, 285

\bibitem[1988]{jansen}
Jansen F., Smith A., Bleeker J.A.M., et al., 1988, ApJ. 331, 949

\bibitem[1994]{jones94}
Jones T.W., Kang H.,  Tregillis I.L., 1994, ApJ 432, 194 

\bibitem[1998]{jones98}
Jones T.W., Rudnick L., Jun B.-I., et al., 1998, PASP 110, 125

\bibitem[1976]{kamper}
Kamper K., van den Bergh S., 1976, ApJS 32, 351


\bibitem[1995]{kassim}
Kassim N.E., Perley R.A., Dwarakanath K.S., Erickson W.C.,
1995, ApJ 455, L59 (KPDE)

\bibitem[1998]{koralesky}
Koralesky B., Rudnick L., Gotthelf E. V.,  Keohane J. W., 
1998, ApJ 505, L27

\bibitem[1979]{klein}
Klein U., Emerson D.T., Haslam C.G.T., Salter C.J., 1979, A\&A
76, 120

\bibitem[1986]{mezger}
Mezger P.G., Tuffs R.J., Chini R., Kreysa E.,  Gemuend H.-P.,
1986, A\&A 167, 145

\bibitem[1995]{reed}
Reed J.E., Hester J.J., Fabian A.C., Winkler P.F., 1995, ApJ 440,
706

\bibitem[1990]{rees}
Rees N., 1990, MNRAS 243, 637

\bibitem[1970]{rosenberg}
Rosenberg I., 1970, MNRAS 151, 109

\bibitem[1999]{osullivan}
O'Sullivan C., Green D. A., 1999, MNRAS 303, 575

\bibitem[1969]{scott69} 
Scott P.F., Shakeshaft J.R., Smith M.A., 1969, Nat 223, 1139 

\bibitem[1975]{scott75}
Scott J.S.,  Chevalier R., 1975, ApJ 197, L5

\bibitem[1996]{the}
The L.-S., Leising M.D., Kurfess J.D., et al., 1996, A\&AS 120, 357

\bibitem[1983]{tuffs83}
Tuffs R.J., 1983, PhD thesis, University of Cambridge

\bibitem[1986]{tuffs86}
Tuffs R.J., 1986, MNRAS 219, 13

\bibitem[1997]{tuffs97}
Tuffs R.J., Drury L., Fischera J., et al., 
1997, Proc. 1-st ISO Workshop on Analytical
Spectroscopy (ESA SP-419), p.177  

\bibitem[1998]{vink}
Vink J., Bloemen H., Kaastra J.S.,  Bleeker J.A.M., 1998, A\&A
339, 201

\bibitem[1990]{woan}
Woan G.,  Duffett-Smith P.L., 1990, MNRAS 243, 87 (WDS90)

\end{thebibliography}
\end{document}